\documentclass[sigconf]{acmart}

\renewcommand\footnotetextcopyrightpermission[1]{} 

\usepackage{booktabs} 
\usepackage{enumerate}
\usepackage{comment}
\usepackage{multirow}
\usepackage{subfigure}
\usepackage{amsmath}
\usepackage[linesnumbered,ruled]{algorithm2e}
\usepackage{algorithmic}

\usepackage{graphicx}
\usepackage{tabularx}
\usepackage{diagbox}
\usepackage{float}
\usepackage{xcolor}
\usepackage{dblfloatfix}
\usepackage{mathtools}
\usepackage{tikz}
\usepackage{cleveref}
\usepackage{balance}

\usepackage{color, colortbl}
\definecolor{Gray}{gray}{0.9}
\definecolor{LightCyan}{rgb}{0.88,1,1}

\newcolumntype{a}{>{\columncolor{Gray}}c}
\newcolumntype{b}{>{\columncolor{LightCyan}}c}

\setcopyright{rightsretained}

\usepackage{pifont}
\newcommand{\cmark}{\ding{52}}%
\begin{document}

\title{Scheduling Beyond CPUs for HPC}

\author{Yuping Fan, Zhiling Lan}
\affiliation{%
  \institution{Illinois Institute of Technology}
  \city{Chicago}
  \state{IL}
}
\email{yfan22@hawk.iit.edu, lan@iit.edu}

\author{Paul Rich, William E. Allcock, Michael E. Papka}
\affiliation{%
  \institution{Argonne National Laboratory}
  \institution{Northern Illinois University}
  \city{Lemont}
  \state{IL}
}
\email{{richp, allcock, papka}@anl.gov}

\author{Brian Austin, David Paul}
\affiliation{%
  \institution{Lawrence Berkeley National Laboratory}
  \city{Berkeley}
  \state{CA}
}
\email{{baustin, dpaul}@lbl.gov}

\begin{abstract}
High performance computing (HPC) is undergoing significant changes. The emerging HPC applications comprise both compute- and data-intensive applications. To meet the intense I/O demand from emerging data-intensive applications, burst buffers are deployed in production systems. Existing HPC schedulers are mainly CPU-centric. The extreme heterogeneity of hardware devices, combined with workload changes, forces the schedulers to consider multiple resources (e.g., burst buffers) beyond CPUs, in decision making. In this study, we present a multi-resource scheduling scheme named BBSched that schedules user jobs based on not only their CPU requirements, but also other schedulable resources such as burst buffer. BBSched formulates the scheduling problem into a multi-objective optimization (MOO) problem and rapidly solves the problem using a multi-objective genetic algorithm. The multiple solutions generated by BBSched enables system managers to explore potential tradeoffs among various resources, and therefore obtains better utilization of all the resources. The trace-driven simulations with real system workloads demonstrate that BBSched improves scheduling performance by up to 41\% compared to existing methods, indicating that \textit{explicitly optimizing multiple resources beyond CPUs is essential for HPC scheduling}.
\end{abstract}

%
%

%

\copyrightyear{2019}
\acmYear{2019}
\setcopyright{usgovmixed}
\acmConference[HPDC '19]{The 28th International Symposium on High-Performance Parallel and Distributed Computing}{June 22--29, 2019}{Phoenix, AZ, USA} 
\acmBooktitle{The 28th International Symposium on High-Performance Parallel and Distributed Computing (HPDC '19), June 22--29, 2019, Phoenix, AZ, USA} 
\acmPrice{15.00}
\acmDOI{10.1145/3307681.3325401} 
\acmISBN{978-1-4503-6670-0/19/06}

\maketitle

\section{Introduction}\label{Introduction}
The exponential growth in computing power has enabled high-performance computing (HPC) systems to attack scientific problems that are much larger and more complex. HPC applications have diverse resource requirements. For them, CPU is not necessarily the main resource determining the required performance, but the allocation with respect to other resources like I/O and network bandwidth becomes more critical \cite{Liu1, Qiao2017,Qiao2018}. A typical example is \textit{data-intensive applications}. These applications have extremely high demand for storage systems. As the growth in computing power continues to outpace the increase in network bandwidth between compute nodes and parallel file system (PFS), PFS fails to rapidly consume bursty data produced by HPC applications. As such, production supercomputers are deployed with \textit{burst buffers} to bridge the performance gap between compute nodes and PFS. Burst buffer is an intermediate storage layer positioned between compute nodes and storage systems. It is typically built from solid-state drive (SSD), offering one to two orders of magnitude higher I/O bandwidth than PFS. Cori \cite{Cori} at National Energy Research Scientific Computing Center (NERSC) and Trinity \cite{Trinity} at Los Alamos National Laboratory (LANL) are deployed with shared burst buffers. 

As burst buffers are incorporated into HPC systems, it is crucial for HPC schedulers to schedule user jobs based on their CPU as well as burst buffer demands. Note that \textit{this study targets at HPC schedulers that are responsible for allocating user jobs onto compute nodes and other system-level schedulable resources, e.g., burst buffers.} The terms CPU and compute node are used interchangeably in this paper. The well-known schedulers in HPC include Slurm, Moab/TORQUE, PBS, and Cobalt \cite{SLURM,Moab,PBS,Cobalt}. Depending on the site mission, HPC facilities deploy different scheduling policies to achieve certain goals \cite{Fan2}. For instance, first come, first served (FCFS) with EASY backfilling is a default scheduling policy deployed at many production systems \cite{Feitelson02}. Despite the use of different scheduling policies, a common goal for HPC scheduling is \textit{to optimize resource utilization}. Existing HPC schedulers are mainly CPU-centric. They often disregard diverse resource requirements and make scheduling decisions solely based on the application's processor footprint \cite{Fan3}. Such a CPU-centric scheduling can easily result in poor application performance and waste of system resources. 

\begin{table*}[h]
\caption{An illustrative example of scheduling multiple resources using different scheduling methods.}  
\centering  
\subtable[Job waiting queue]{  
\resizebox{0.2\linewidth}{!}{
\begin{tabular}{l|l|l}
\hline
\rowcolor{LightCyan}
Job & Nodes & Burst Buffers (TB) \\ \hline
\rowcolor{Gray}
J1  & 80   & 20                 \\ \hline
\rowcolor{LightCyan}
J2  & 10   & 85                \\ \hline
\rowcolor{Gray}
J3  & 40   & 5                \\ \hline
\rowcolor{LightCyan}
J4  & 10   & 0                \\ \hline
\rowcolor{Gray}
J5  & 20   & 0                \\ \hline
\end{tabular}
}
\label{example_job_requests}  
}  
\subtable[The scheduling decisions made by different scheduling methods]{          
\resizebox{0.76\linewidth}{!}{
\begin{tabular}{|c|c|c|c||c|c|c|c|b|}
\hline
Solution & \begin{tabular}[c]{@{}c@{}}Selected\\ Jobs\end{tabular} & \begin{tabular}[c]{@{}c@{}}Node  \\ Utilization\end{tabular} & \begin{tabular}[c]{@{}c@{}}Burst Buffer \\ Utilization\end{tabular} & \begin{tabular}[c]{@{}c@{}}Naive \\ Method\end{tabular} & \begin{tabular}[c]{@{}c@{}}Constrained \\ Method\end{tabular} & \begin{tabular}[c]{@{}c@{}}Weighted\\  Method\end{tabular} & \begin{tabular}[c]{@{}c@{}}Bin\\  Packing\end{tabular} &\begin{tabular}[c]{@{}c@{}}Pareto\\  Set\end{tabular}\\ \hline
1        & J1, J4                                                  & 90\%                                                        & 20\%                                                  & \cmark   &       &             &                 &                      \\ \hline
2        & J1, J5                                                  & 100\%                                                        & 20\%                                                             &      &                        \cmark                                 &              \cmark               &              \cmark              &         \cmark             \\ \hline
3        & J2, J3, J4, J5                                                  & 80\%                                                        & 90\%                                                             &                                                      &         &                    &               &          \cmark         \\ \hline
\end{tabular}
}
 \label{example_results}  
}  
\end{table*}   

Slurm is a well-known scheduler that supports burst buffer scheduling \cite{SLURMBB}. Slurm allocates the jobs from the waiting queue in sequence until either CPU or burst buffer is exhausted. We denote it as \textit{naive method} in this study. This approach has a limited efficiency: the depletion of one resource can prevent the queued jobs from allocation, causing under-utilization of the other resource. Two optimization approaches for solving multi-resource scheduling problems may be applicable for co-scheduling CPU and burst buffer.
One approach is to optimize utilization of one resource and treat other resources as constraints (denoted as \textit{constrained method}) \cite{xu01,Wallace,Rao}. Another approach is to combine utilizations of multiple resources into one objective by a weighted sum (denoted as \textit{weighted method}) \cite{Ren,Hung,Jakob}. 
Both optimization methods convert a multi-resource scheduling problem into a single-objective optimization problem. Such a conversion leads to the loss of a prominent characteristic of multi-resource scheduling, i.e., trade-offs between competing resources. \textit{Bin packing} is discussed in the literature to improve resource utilization for cluster scheduling \cite{Grandl,NoroozOliaee}. It models machines as bins and tasks as balls. Balls of different volumes are packed into bins of certain capacities iteratively and the goal is to minimize the number of bins used. This simple heuristic selects jobs in a one-by-one manner, which may miss the best job combination that maximizes resource utilization. 

\textbf{An Illustrative Example:} Here we give a simple example to show the limitations of the existing scheduling methods for scheduling CPU and burst buffer on HPC. Consider a system with 100 nodes and 100TB of burst buffers. Five jobs are in the queue, each having different resource demands as shown in Table \ref{example_job_requests}.

Table \ref{example_results} compares the scheduling results of different methods. A naive method selects J1 and backfills J4 (explained in \Cref{HPC Job Scheduling}) for execution, resulting in node utilization of 90\% and burst buffer utilization of 20\%. Such a scheduling wastes 80TB of burst buffers and prevents other jobs from being scheduled. A constrained method may optimize node utilization under the constraint of the burst buffers. A weighted method may use a linear combination of node utilization with 80\% weight and burst buffer utilization with 20\% weight as the objective. A bin packing method may pick jobs with the maximum dot product between the demands of the job and the remaining amount of resource iteratively. The constrained, the weighted and the bin packing methods select J1 and J5 for execution, achieving node utilization of 100\% and burst buffer utilization of 20\% (Solution 2). While these methods improve node utilization, they still leave 80\% of the burst buffers wasted.

All these methods overlook an alternative solution: the selection of J2-J5 for resource allocation by skipping J1 (Solution 3). Solution 3 achieves significantly higher burst buffer utilization, while slightly lowering node utilization as compared to Solution 2. This simple example highlights the importance of identifying a \textit{Pareto set}\footnote{Pareto set is a set of non-dominated solutions, being chosen as optimal, if no objective can be improved without sacrificing at least one other objective \cite{Reddy}. For the example shown above, the Pareto set contains Solution 2 and 3.}, each solution in the Pareto set representing a tradeoff among different objectives (i.e., the selection of different resources). 

In this study, we present a multi-resource scheduling scheme denoted as \textit{BBSched} that allocates multiple resources to user jobs based on their resource demands.  \textit{Distinguishing from existing methods, BBSched aims to optimize the utilization of multiple resources by providing a Pareto set for decision making.} There are three \textit{key obstacles} to overcome in the design of BBSched. First, the design has to be practical in the sense that it can make rapid scheduling decisions. Current HPC systems typically require a scheduler to respond in 15-30 seconds \cite{xu01,Verma}. Second, an efficient design has to improve system-related performance with minimal impact on site policies. Finally, HPC is dynamically evolving such that systems are constantly expanded with new resources. Hence, the scheduler is expected to be extensible to embrace emerging resources.

To tackle the above obstacles, several techniques are explored for the BBSched design. First, BBsched is developed as a plugin to existing HPC schedulers (denoted as \textit{base schedulers}) to preserve job priority according to a site's policy. Unlike the traditional one-by-one job selection used in the conventional scheduling, BBSched leverages a window-based scheduling approach to dispatch a set of jobs from the front of job waiting queue. Such a window-based design aims to maintain the job ordering given by the base scheduler. Second, jobs are selected from the window for resource allocation, with the objective to optimize resource utilization. We formulate the multi-resource scheduling problem into a multi-objective optimization (MOO) problem. Contrary to a single-objective optimization, our MOO formulation simultaneously optimizes the utilizations of multiple resources and returns a Pareto set. The Pareto set provides a set of optimal solutions which enables system managers to make a scheduling decision  by considering the tradeoffs among different optimal solutions. Considering that MOO is NP-hard \cite{Bringmann}, we explore a genetic algorithm as the MOO solver for meeting the rigid time requirement. 

We evaluate BBSched by means of extensive trace-based simulations with real workload traces collected from Cori at NERSC and Theta \cite{Theta} at Argonne Leadership Computing Facility
(ALCF). Additionally, we generate a series of workloads based on the real traces to stress various resource usages. The goal is to extensively evaluate BBSched under various scenarios, especially under the cases of resource confliction and saturation. A series of experiments are conducted to compare BBSched with existing methods (naive, constrained, weighted, and bin packing methods) on scheduling CPU and burst buffer. The results show that BBSched is capable of improving resource utilization by up to 20\% and reducing average job wait time by up to 41\%. 

Furthermore, we present a case study to show BBSched can be easily extended to schedule additional resources beyond CPUs and burst buffer. The preliminary results clearly indicate that BBSched outperforms existing methods in terms of both system-level and user-level scheduling metrics. \textit{This demonstrates that explicitly optimizing all resources is crucial to multi-resource scheduling.} 

The remainder of this paper is organized as follows. We start by introducing background and related work in Section \ref{Background}, including HPC scheduling, burst buffer, and multi-resource scheduling. Section \ref{Methodology} describes our design. The experimental results of scheduling CPU and burst buffer are presented in Section \ref{Experiments}. A case study of incorporating more resources in BBSched is examined in Section \ref{Extensibility}. Finally, we conclude the paper in Section \ref{Conclusion}.

\section{Background and Related Work}\label{Background}
\subsection{HPC Scheduling}\label{HPC Job Scheduling}
System-level HPC scheduling, also known as batch scheduling, is responsible for assigning jobs to resources according to site policies and resource availability. It targets on scheduling compute nodes along with other system-level resources. This is different from core-level application scheduling or task scheduling that is typically handled by operating systems. Well-known schedulers in HPC include Slurm, Moab/TORQUE, PBS, and Cobalt \cite{SLURM,Moab,PBS,Cobalt}. When submitting a job, a user is required to provide two pieces of information: resources required by the job and runtime estimate \cite{Fan1}. The jobs are stored and sorted in the waiting queue based on a site's policy. In the past, a number of scheduling policies have been proposed, and one of the widely used policies is FCFS, which sorts the jobs in the order of their arrivals. At ALCF, to support the mission of running large-scale capability jobs, a utility-based scheduling policy, named WFP, is deployed which periodically calculates a priority increment for each waiting job \cite{Fan2,Yu2018}. EASY backfilling is a commonly used strategy to enhance system utilization, where subsequent jobs are allowed to skip ahead under the condition that they do not delay the job at the head of the queue \cite{Feitelson02}. 

In this study, we denote the above schedulers that enforce job priority according to a site's policy as base schedulers. BBSched can be used along with these base schedulers, for optimizing utilization of multiple resources, without unnecessary impact of job priority posed by the base scheduler.

\subsection{Burst Buffer}
HPC systems are facing the challenge of ever-growing gap between compute power and I/O performance. Bridging this gap becomes increasingly critical with the increasing data-intensive applications. I/O behavior of data-intensive applications is characterized by intense bursts of data access \cite{Liu1}. Burst buffers, an intermediate storage layer between compute nodes and PFS, are designed to absorb bursty I/O data effectively. They are typically built from SSD, providing significantly higher bandwidth and lower latency than PFS. A burst buffer can be either attached to compute nodes as a local resource or configured as a global resource shared by compute nodes. 
Cori at NERSC and Trinity at LANL adopt shared burst buffers; Theta at ALCF and Summit \cite{Summit} at Oak Ridge Leadership Computing Facility (OLCF) are equipped with local SSDs.

Existing studies on burst buffer scheduling are mainly at application level or at I/O server level. Little work has been done at system scheduling level. Slurm supports co-scheduling of CPU and burst buffer; however, it lacks optimization for CPU and burst buffer. This study will address the co-scheduling of CPU and burst buffer with the objective of optimizing the utilization of both resources.
\subsection{Multi-Resource Scheduling}\label{Multi-objective optimization}
Considerable research exists in exploring optimization methods for multi-resource scheduling. 
Constrained optimization is a common optimization method. For example, Wallace et al. addressed a power-aware scheduling problem by optimizing node utilization with a power limit \cite{Wallace}; Rao et al. examined the problem of reducing the total electricity cost while guaranteeing the quality of service in datacenters \cite{Rao}; 
Xu et al. presented an energy-aware scheduling framework which maximizes power consumption at off-peak time with the constraint of nodes \cite{xu01}. Weighted sum is another widely adopted optimization method. For example, Ren et al. converted an energy-aware cluster scheduling problem to optimization of the weighted sum of energy cost and fairness \cite{Ren}; Huang et al. treated the problem of multi-resource allocation in geo-distributed clusters as the minimization of the sum of the time spent on network transfer and computation \cite{Hung}; 
Jakob et al. formulated a workflow scheduling problem in grid into optimizing a weighted sum of execution time, costs, and resource utilizations \cite{Jakob}. The above studies basically convert multi-objective optimization problem into a single objective optimization problem through a weighted combination or a constrained approach. In contrast to these studies, we formulate the problem as a MOO problem and use a multi-objective genetic algorithm for solving the MOO problem. Unlike the constrained or the weighted optimization that only provides a single solution optimized for a first-class objective or a weighted sum of the objectives, our method is capable of optimizing multiple objectives by providing a Pareto set for decision making.

Multi-resource cluster scheduling is an active research area that has received much attention in the past years. Cluster schedulers typically emphasize fair resource allocation 
\cite{DRF,Wang,Chowdhury,Grandl,Joe}, which distracts considerably from the goal of HPC scheduling, i.e., high resource utilization. For example, dominant resource fairness (DRF) allocates multiple resources satisfying strategy-proof, envy-freeness, sharing incentive and pareto-efficiency \cite{DRF}.
Pareto efficiency is defined as increasing the allocation of a user should not decrease the allocation of at least another user \cite{DRF}, which is different from the Pareto set targeted in this work.
In practice, production cluster schedulers also prefer fairness as their primary scheduling goal \cite{Apollo,Mesos,Omega,YARN}. However, fairness and utilization are conflicting goals and aggressively using fair sharing can hurt cluster utilization \cite{Altruistic}. In contrast, HPC systems are designed to run large jobs and prefer users running large jobs \cite{Fan2}. Therefore, fair sharing is not a concern in HPC scheduling. Some other cluster schedulers focus on individual job performance, e.g., lowering job completion times \cite{NoroozOliaee} and responding timely to latency-critical services \cite{Zhang01}. Due to the heterogeneous nature of cluster machines, job completion times vary when jobs run on different machines. This assumption does not hold on HPC systems. Moreover, to ensure timely response to latency-critical services, clusters have to reserve resources for unexpected load spikes, leading to low resource utilization \cite{Zhang01}.

A few studies in multi-resource cluster scheduling evaluated the negative impact of fairness on utilization and made tradeoffs between fairness and utilization \cite{Grandl,Altruistic}. Owing to substantial job arrival rate, cluster schedulers have to respond in seconds or shorter. Hence, they seek quick but greedy methods for scheduling. For instance, Grandl et al. have explored the potential of using multi-dimensional bin packing to improve resource utilization for multi-resource clusters \cite{Grandl}. Although bin packing algorithms are capable of making timely scheduling decisions, they allocate jobs in a one-by-one manner based on individual job information and therefore lead to less desirable performance as compared to the MOO approach adopted in this study which jointly considers resource requirements of multiple jobs. Altruistic scheduling improves resource utilization as well as meets fair-share guarantees by redistributing leftover resources, i.e., fractions of allocated resources \cite{Altruistic}. This is inapplicable to HPC scheduling because the sizes of HPC jobs are fixed throughout their execution.

\section{Methodology}\label{Methodology}

\begin{figure}[h]
\centering
\includegraphics[scale=0.33]{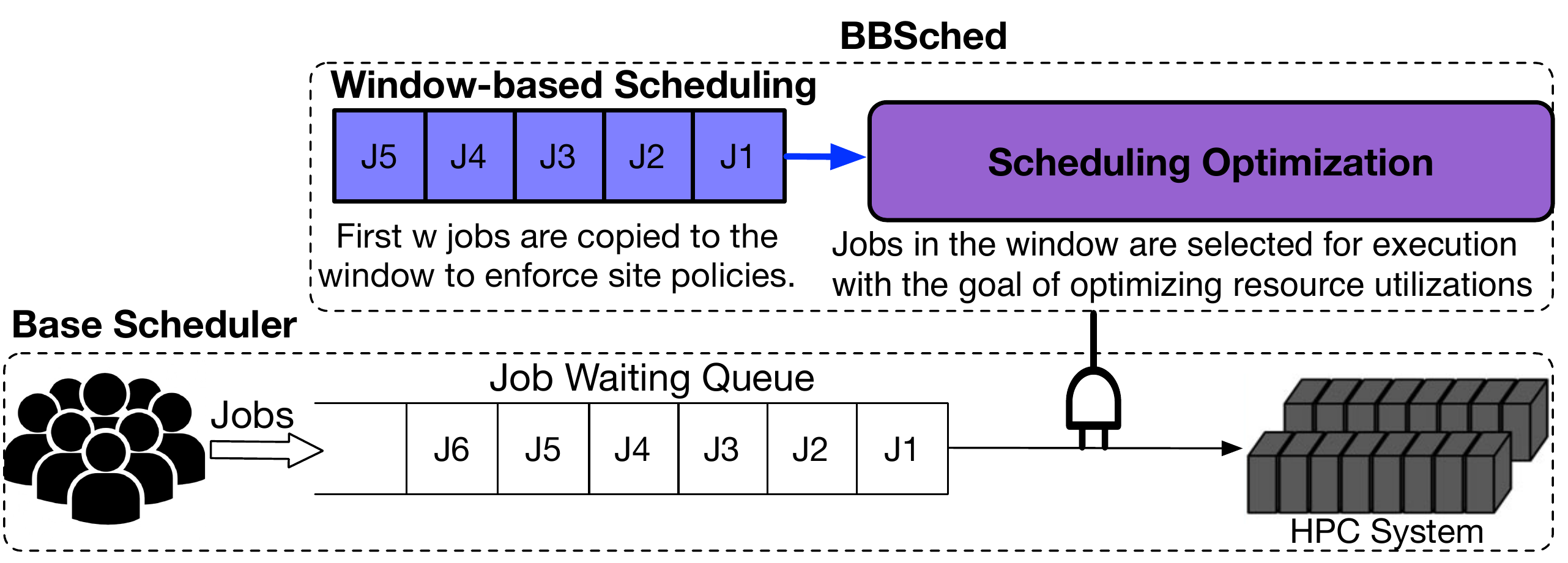}
\vspace*{-0.55cm}
\caption{The overview of BBSched.}
\label{system_overview_fig}
\end{figure}

In this section, we present BBSched for HPC scheduling under multiple resource constraints. As shown in Figure \ref{system_overview_fig}, BBSched is built as a plug-in to a base scheduler which enforces job priority according to a site's policy. BBSched consists of two components: \textit{a window-based scheduling and a scheduling optimization scheme}. BBSched begins with a window-based scheduling with the aim to balance scheduling performance and enforcing site policies (\Cref{Window-Based Mechanism}). The jobs in the window are selected to execute for optimizing resource utilizations (\Cref{Burst Buffer Aware Scheduler}). The optimization process first formulates the multi-resource scheduling problem into a multi-objective optimization (MOO) problem, which optimizes node utilization and burst buffer utilization (\Cref {MOO Formulation}). Given that this MOO problem is NP-hard, we then develop an efficient meta-heuristic for problem-solving (\Cref {MOO Solver}). We discuss how to select the parameters in the solver in \Cref {Parameter Selection}. The final solution is chosen from multiple solutions by considering tradeoffs among multiple resource usages (\Cref {Decision Making}). Finally, we analyze the computational complexity of BBSched in \Cref{Complexity Analysis}.

\subsection{Window-Based Scheduling}\label{Window-Based Mechanism}
Rather than allocating jobs one by one from the front of the waiting queue, we adopt the window-based scheduling which allocates multiple jobs from a window at the front of the waiting queue \cite{xu01}. In doing so, we balance the goals of optimizing system metrics and enforcing site policies.
Note that jobs with dependencies are allowed to enter the window only if all the dependencies have been completed. This restriction keeps dependent jobs in order and preserves the priority of jobs with dependencies. 
Window size is a tunable parameter. The selection of window size is dependent on site policy and workload characteristics. A larger window size means more jobs are considered for optimization; however, it also means less preservation of the original job order. System managers may choose a window size according to their preference for more optimization or more preservation of job order. In addition, the window size could be dynamically adjusted in response to system status. Job queue length often changes. For instance, it is typically longer during workdays and is shorter during weekends. In this study, we use a static window size.

An issue with the window-based scheduling is \textit{job starvation}, meaning that a job may stay in the window without being selected to execute. To prevent job starvation, we define an upper bound for the number of iterations that a job can stay in the window. Once a job passes the bound (e.g., 50), it must be selected to run.

\subsection{Scheduling Optimization}\label{Burst Buffer Aware Scheduler}
\subsubsection{\textbf{Multi-Objective Optimization (MOO) Formulation:}}\label{MOO Formulation}
The optimization process determines which jobs are selected from the window to execute so that node utilization and burst buffer utilization are maximized. To achieve system performance goal, we formulate the multi-resource scheduling problem into the following MOO problem.

Suppose a system has $N$ nodes and burst buffers of total $B$ GB. Assume that upon a scheduling invocation, the amounts of nodes and burst buffers being used are $N_{used}$ and $B_{used}$ respectively. Suppose $J = \{J_1,\dots,J_w\}$ is a set of $w$ jobs in the scheduling window: job $J_i$ requiring $n_i$ nodes and $b_i$ GB of burst buffers. 

The scheduling problem can be transformed into the following MOO: \textit{to determine a finite set of Pareto solutions $\boldsymbol{X}$}; each Pareto solution $\boldsymbol{x} \in \boldsymbol{X}$ is represented by a binary vector $\boldsymbol{x} = [x_1, \dots,x_w]$, such that $x_i=1$ if $J_i$ is selected to execute and $x_i=0$ otherwise. A Pareto solution optimizes the following two objectives:
\begin{enumerate}
\item maximize node utilization: $f_1(\boldsymbol{x}) = \sum \limits_{i=1}^w n_i \times x_i$
\vspace*{-0.2cm}
\item maximize burst buffer utilization: $f_2(\boldsymbol{x}) = \sum \limits_{i=1}^w b_i \times x_i$
\vspace*{-0.2cm}
\end{enumerate} 
Formally, the problem can be formulated as:
\begin{align}
\text{max}  &  \quad  (f_1(\boldsymbol{x}),f_2(\boldsymbol{x})) \notag\\
\text{s.t.} &  \quad   \sum \limits_{i=1}^w n_i \times x_i \le N-N_{used},   \quad x_i \in \{0, 1\} \notag\\
&  \quad  \sum \limits_{i=1}^w b_i \times x_i \le B-B_{used},  \quad x_i \in \{0, 1\} \notag
\end{align}
where the constraints guarantee that assigned resources do not exceed available nodes and burst buffers in a system. 

\begin{figure}[h]
\centering
\includegraphics[width=0.4\textwidth,height=1.5in]{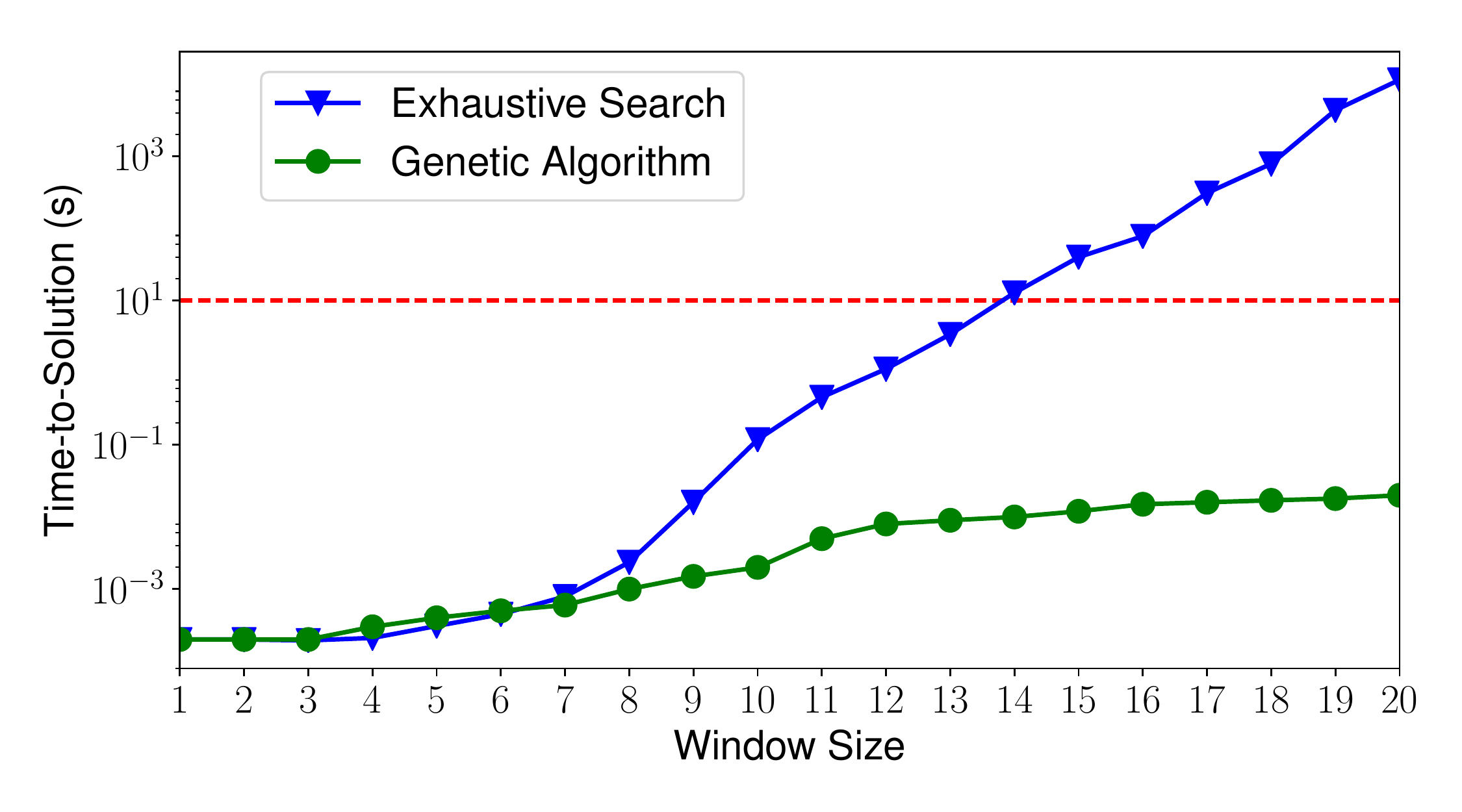}
\caption{Impact of window sizes on average solution time. Figure \ref{exhaustive_genetic} and \ref{genetic_alg}  were conducted with first 1000 jobs from a Theta workload (see Table \ref{Cori_Theta_comp}). Solutions above the red dash line do not meet the time requirement of HPC scheduling.}
\label{exhaustive_genetic}
\end{figure}

\subsubsection{\textbf{MOO Solver:}}\label{MOO Solver}
The above MOO problem is NP-hard. To find all solutions, one has to exhaustively examine $2^w$ possible solutions and compare them to determine a Pareto set. As the window size $w$ increases, the number of possible solutions as well as the time-to-solution increases exponentially (see Figure \ref{exhaustive_genetic}). Current HPC systems typically require a scheduler to respond in 15-30 seconds \cite{xu01,Verma}. To achieve fast decision making, we need a rapid solver to solve the MOO problem. In this study, we explore a multi-objective genetic algorithm \cite{Konak} to solve the MOO problem. \textit{This genetic-based algorithm approximates the true Pareto set iteratively,} so it requires much less time. It can be accelerated by leveraging parallel processing \cite{Deb}. 
A genetic algorithm attempts to mimic natural selection: the population size is a constant $P$; weak chromosomes are extinct by natural selection, while strong chromosomes survive and pass their genes to future generations. 

\begin{figure}[!h]
\centering
\includegraphics[scale=0.24]{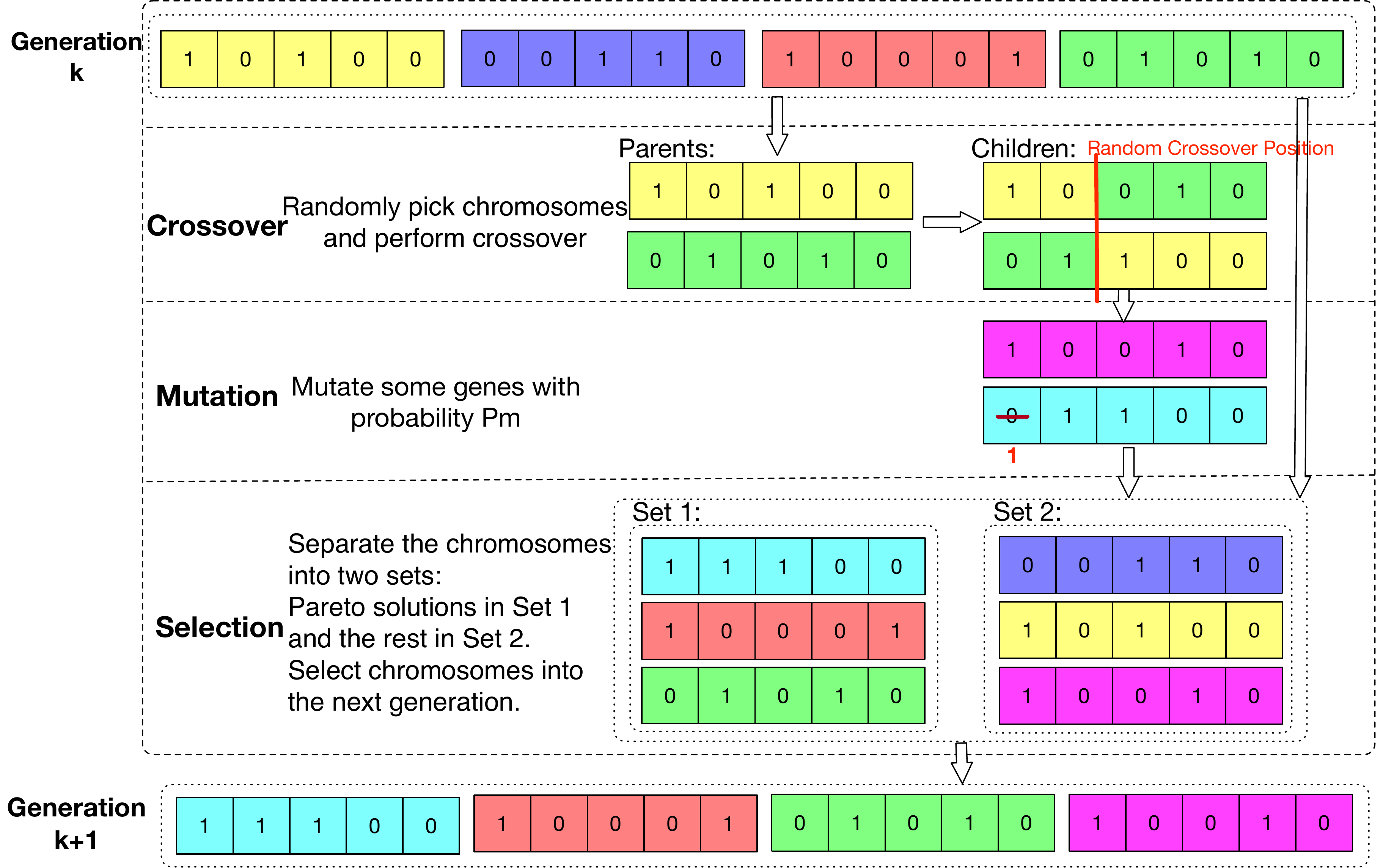}
\caption{An example of the evolution process to solve the MOO problem. MOO solver maintains a population of can- didate solutions (4 chromosomes). A chromosome consists of 5 genes, where each gene represents the selection of the job at a specific location in the window and encodes as a bi- nary number: 1 (selected) or 0 (not selected).}
\label{problem_for_fig4}
\end{figure}

Figure \ref{problem_for_fig4} illustrates the process of solving the MOO problem. 
The first generation is initialized randomly. Generations are evolved iteratively via crossover, mutation, and selection operations.
The crossover operation generates two children by randomly selecting two parents from the previous generation and swapping genes of parents at a random position.
The mutation is used to introduce diversity and to prevent our solver from trapping in local optima. With a low probability $p_m$, the genes of the children are bit flipped. A high mutation rate leads to a random search and results in poor solutions.
The selection operation constructs a new generation by carrying over good chromosomes to the next generation. Specifically, we separate the chromosomes into two sets: Pareto solutions in Set 1 and the rest in Set 2. A solution is chosen as a Pareto solution, if improving one of its objectives would deteriorate at least one other objective. If Set 1 has less than $P$ chromosomes, all the chromosomes in Set 1 are passed to the next generation and then chromosomes in Set 2 (newer chromosomes have higher priorities). If Set 1 has more than $P$ chromosomes, we select those with newer ages. Upon an evolution to new generation, the ages of chromosomes are increased by 1.

When the number of generations reaches a predefined threshold $G$, the above iterative process stops and the chromosomes in Set 1 in the final generation form a Pareto set.

\subsubsection{\textbf{Parameter Selection:}}\label{Parameter Selection}
The solver contains three parameters: number of generations ($G$), population size ($P$), and mutation probability ($p_m$). According to the literature, $p_m$ is normally very low, i.e., less than 0.1\% \cite{Haupt,Eiben}. In our experiments, varying $p_m$ has negligible effect on approximation accuracy. 
A larger value of $G$ and $P$ means exploring a larger search space for optimization but more time to solve the problem. Therefore, the selection of $G$ and $P$ is a trade-off between performance and time-to-solution.
The widely adopted metric \textit{generational distance (GD)} is used to measure the accuracy of the solutions. GD is defined as:
\begin{equation}
GD(S) = avg_{u \in S}(min_{v \in S^*}(dist(u,v))) \notag
\end{equation}
where $S$ is the solution set obtained by our MOO solver; $S^*$ is the true Pareto set.
$GD$ computes the average distances between our solution and its nearest true Pareto solution. The smaller the GD is, the better the performance is.

\begin{figure}[h]
\centering
\includegraphics[scale=0.3]{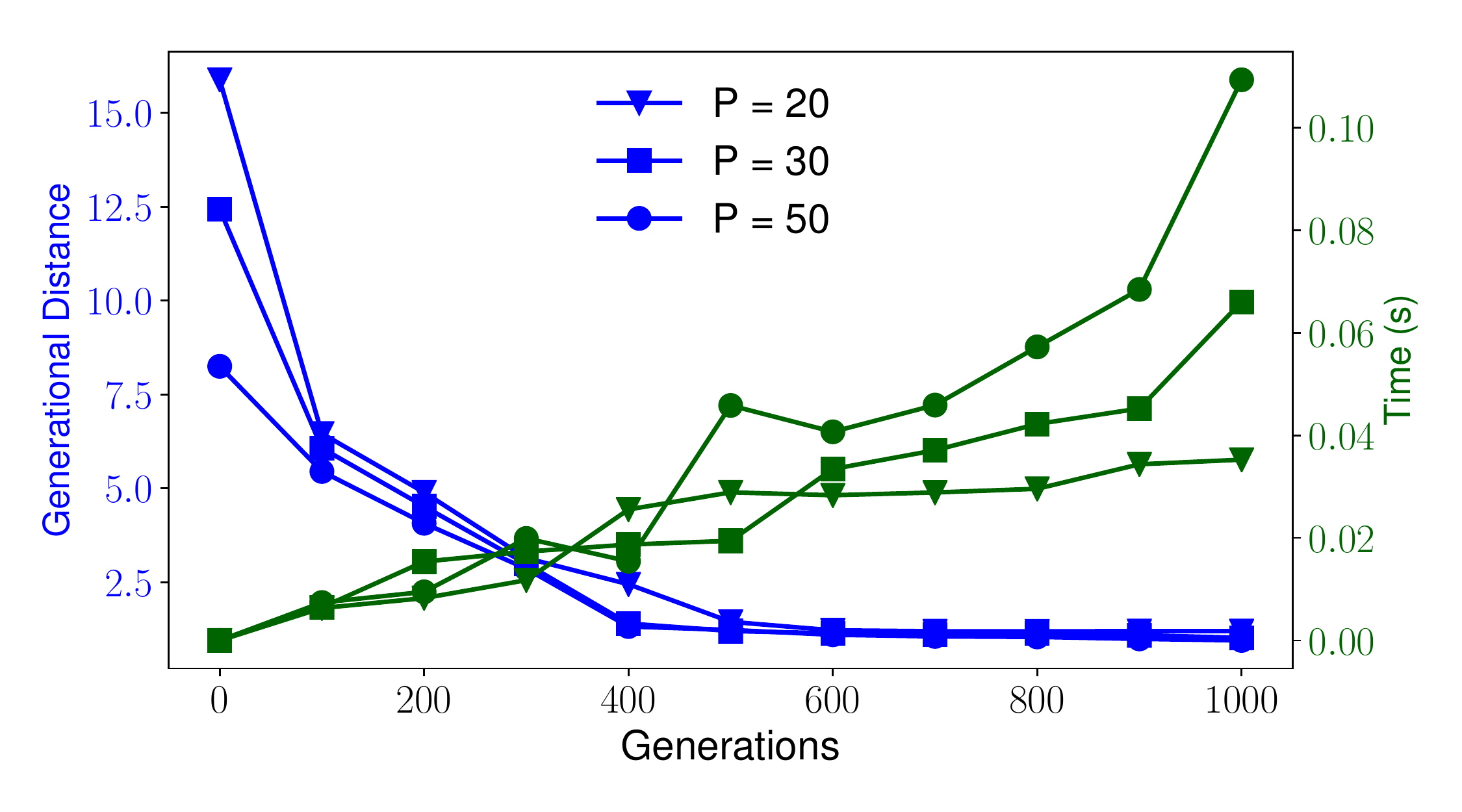}
\caption{Impact of varying $G$ and $P$ parameters.}
\label{genetic_alg}
\end{figure}

Figure \ref{genetic_alg} illustrates an example of the $GD$ value and the time-to-solution as $G$ and $P$ vary. Clearly, as $G$ increases, $GD$ decreases and time-to-solution increases. For $GD$, the most significant improvement is between 0 and 500 generations and the improvement slows down after 500 generations. We also notice that increasing $P$ leads to a decrease in $GD$ and increase in time. This example suggests that our MOO solver is capable of achieving accurate approximations with minimal overhead (less than 0.2 second).  For the workloads investigated in \Cref{HPC Workload Traces}, setting $G=500$ and $P=20$ offers the best tradeoff between accuracy and time-to-solution. 

\subsubsection{\textbf{Decision Making:}}\label{Decision Making}
The output of the solver is a Pareto set, and a decision maker needs to select one preferred solution. Different HPC facilities may have different site policies and scheduling priorities. \textit{System managers may use a site-specific metric for selecting a preferred solution out of the Pareto set.} In this study, we use the following rule. It first chooses the solution that maximizes node utilization, and in case of multiple such solutions, it selects the solution containing the jobs at the front of the window so as to preserve the original job order. Next, it compares the solution with other Pareto solutions for tradeoff analysis. The preferred solution is replaced by another solution if the improvement on the burst buffer utilization is more than 2x of the loss of the node utilization. If more than one such solutions exist, the solution with the maximum improvement is chosen. 

The decision making may be \textit{adaptive}, such that system managers dynamically adjust their selection policy according to scheduling performance and user response. This adaptive decision making is out of the scope of this work and is a topic of our future work.

\subsection{Complexity Analysis}\label{Complexity Analysis}
The window-based mechanism takes a constant time $O(1)$. The scheduling optimization iterates $G$ times. In each iteration, crossover, mutation, and selection are operated on $P$ chromosomes. In total, the optimization requires $O(G \times P)$ operations in the worst cases. Therefore, the time complexity of our design is $O(G \times P)$. This cost can be further lowered via parallel processing of the MOO.

\section{Evaluation}\label{Experiments}
In this section, we evaluate BBSched through extensive trace-based simulation using real workload traces collected from production systems. We describe the two real workload traces collected from Cori and Theta and the eight synthetic workloads with various burst buffer demands derived from the real traces (\Cref{HPC Workload Traces}). We then list the system- and user-centric metrics for scheduling evaluation (\Cref{Evaluation Metrics}) and the multi-resource scheduling methods for comparison (\Cref{Scheduling Methods}). Finally, we quantify BBSched's performance improvements over existing methods (\Cref{Results}).

\begin{table}[htp]
\centering
\caption{Overview of Cori and Theta workloads.}
\label{Cori_Theta_comp}
\resizebox{\linewidth}{!}{
\begin{tabular}{|l|l|l|}

\hline
\rowcolor{LightCyan}
                              & Cori                                                                                            & Theta                 \\ \hline
\rowcolor{Gray}
Location                      & NERSC                                                                                           & ALCF                  \\ \hline
\rowcolor{LightCyan}
Scheduler             & Slurm                                                                                            & Cobalt                   \\ \hline
\rowcolor{Gray}
System Types                     & Capacity computing                                                                                     & Capability computing            \\ \hline
\rowcolor{LightCyan}
Compute Nodes                         & \begin{tabular}[c]{@{}l@{}}12,076\\ (2,388 Haswell; 9,688 KNL)\end{tabular} &  \begin{tabular}[c]{@{}l@{}}4,392\\ (4,392 KNL)\end{tabular}                 \\ \hline
\rowcolor{Gray}
Aggregated Memory             & 1,304.5TB                                                                                        & 913.5TB               \\ \hline
\rowcolor{LightCyan}
Shared Burst Buffer                  & 1.8PB                                                                                           & 1.26PB   (projected)  \\ \hline
\rowcolor{Gray}
Trace Period                  & Apr. 2018 - Jul. 2018                                                                           & Jan. 2018 - May. 2018 \\ \hline
\rowcolor{LightCyan}
Number of Jobs & 2,607,054                                                                                       & 70,507                \\ \hline
\rowcolor{Gray}
BB Data Source                  & Slurm log                                                                         & Darshan log \\ \hline
\rowcolor{LightCyan}
BB Range & [1GB, 165TB]                                                                                      & [1GB, 285TB]                \\ \hline
\end{tabular}
}
\end{table}

\subsection{Workload Traces}\label{HPC Workload Traces}
Table \ref{Cori_Theta_comp} summarizes the real workload traces used in this study. They represent two typical HPC workloads: one for \textit{capacity computing} and the other for \textit{capability computing}. Note that both traces do not include job dependency information and therefore we suppose all jobs are independent in our experiment. The first workload is a four-month job log on Cori from April to July in 2018. Cori is deployed with a 1.8PB Cray Data Warp burst buffers. It uses Slurm for job scheduling. The Slurm log records a number of information per user job which include the requested number of nodes, the requested burst buffer size, job runtime estimate, job submit time, etc. Besides few extremely large requests (165TB), the burst buffer requests are in the range of [1GB, 65TB]. Among all the jobs, 0.618\% of jobs request burst buffers and 0.204\% of jobs request more than 1TB burst buffers. Burst buffers on Cori can be either assigned to individual jobs or assigned to users as persistent reservations. One-third of burst buffers on Cori are reserved persistently and their lifetimes are independent of jobs.

The second workload is a half-year job log on Theta from January to May in 2018. While Theta is currently not deployed with any shared burst buffer, we enhance the trace with burst buffer requests by assuming there was a shared burst buffer of 2.16PB. This assumption is based on the ratio of aggregated memory to the total burst buffer volume on Cori and the aggregated memory on Theta. The trace from Theta contains all the necessary information except for the requested burst buffer size. To address this problem, we use a corresponding \textit{Darshan} \cite{Darshan} trace to extract the amount of data moved between PFS and nodes and consider this amount to be the potential burst buffer requests.
In the half-year workload, 40\% of jobs on Theta have Darshan I/O recording. 17.18\% of the jobs have more than 1GB data transferred and we set the amount of transferred data as the corresponding job's burst buffer request. The requested burst buffer sizes are in the range of [1GB, 285TB].

\begin{figure}[h]
  \centering
  \subfigure{\includegraphics[width=0.23\textwidth,height = 3.2in]{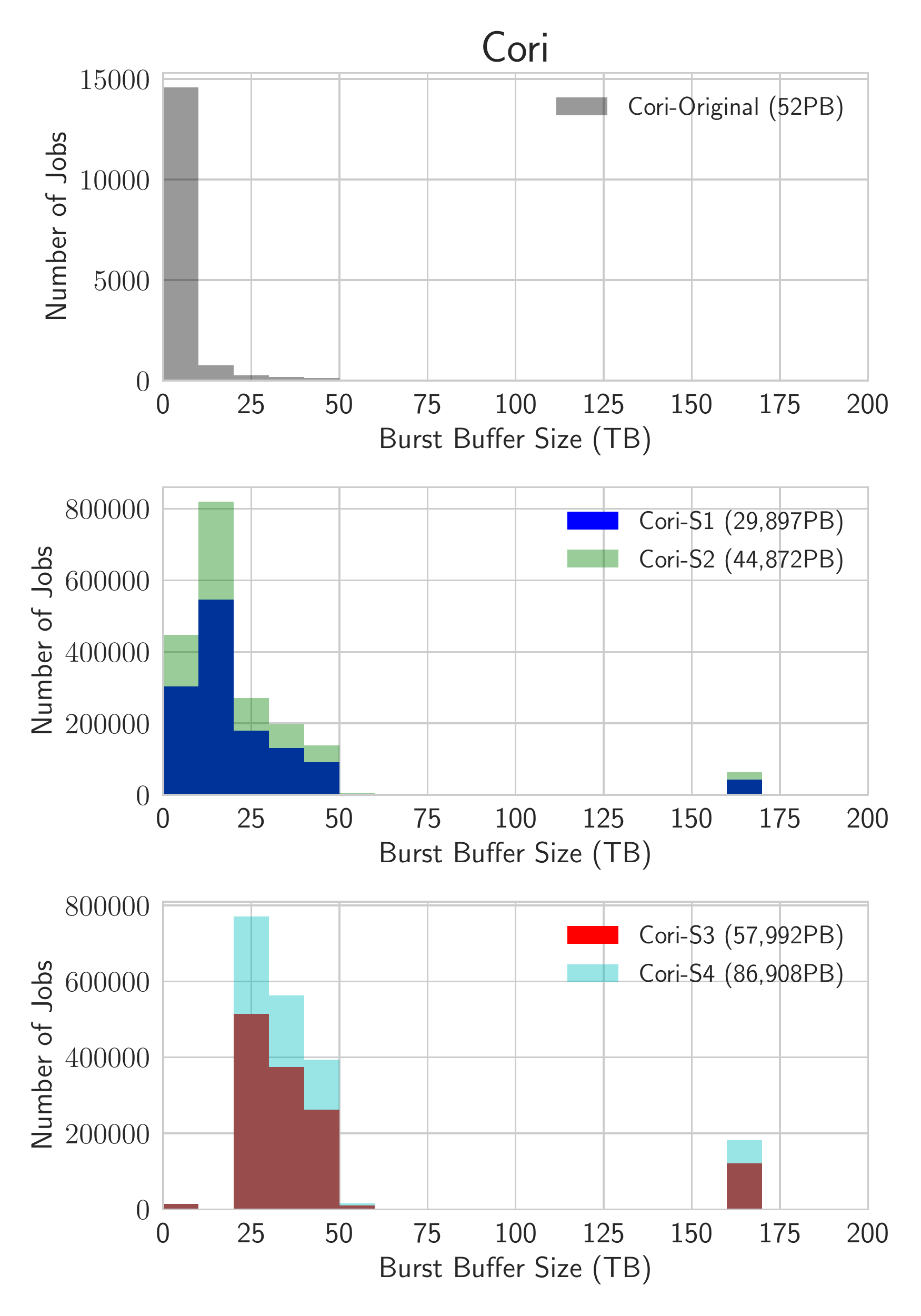}}
  \subfigure{\includegraphics[width=0.23\textwidth,height = 3.2in]{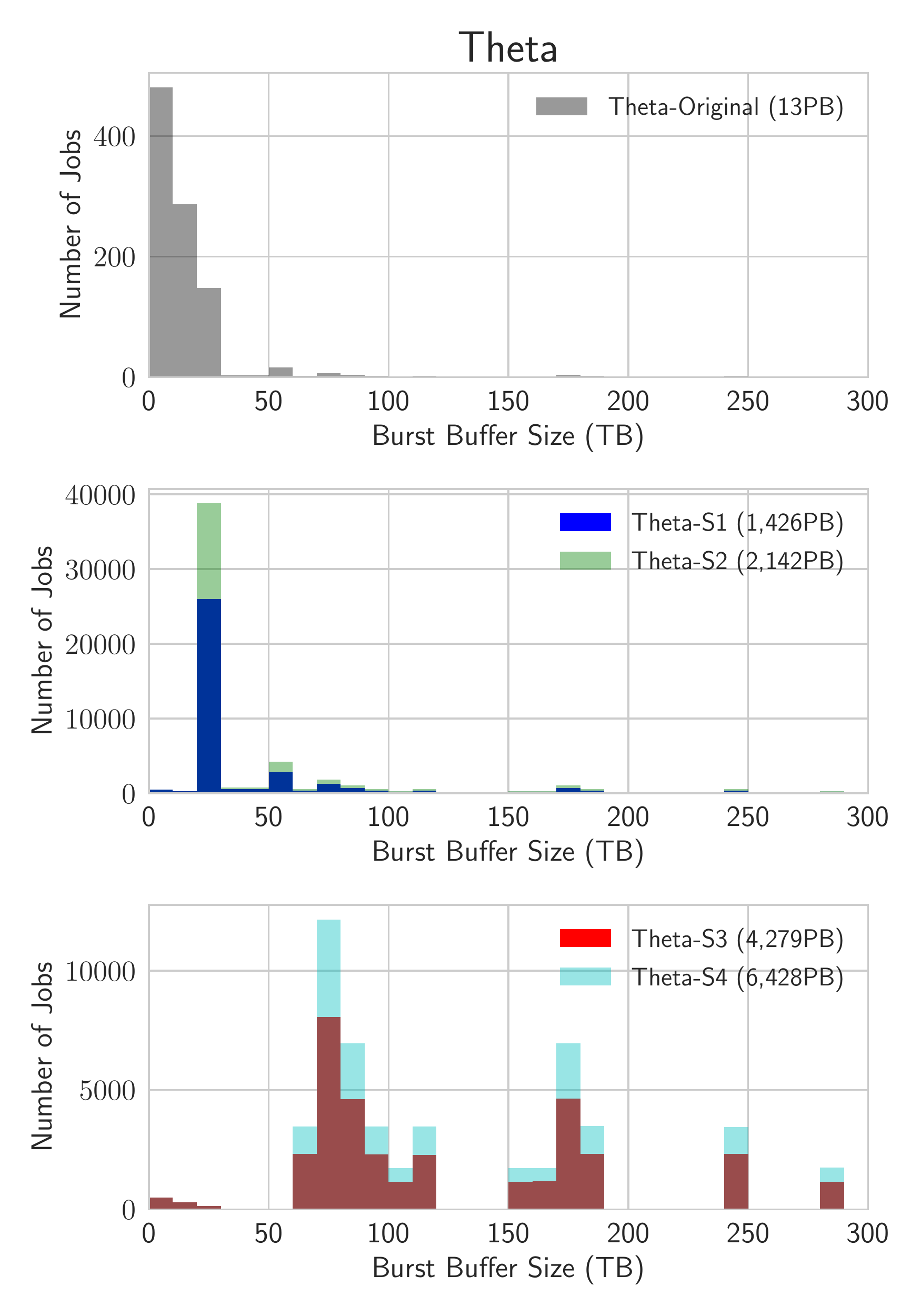}}
  \caption{Histograms of burst buffer distributions on Cori (left) and Theta (right). The bin size is 10TB. The number in the parenthesis is the aggregated volume of requested burst buffers in each workload.}
  \label{workloads_pdf}
\end{figure}

\textit{One issue with both traces is that burst buffers are not heavily used.} There are two possible reasons. First, some jobs have burst buffer demands but are currently not recorded in the system logs. Second, burst buffer is a relatively new resource for users. As users are getting more exposure to it, we expect significant increases in requests for burst buffers.
Having this log limitation in mind, in addition to the two original workloads, we create eight synthetic workloads, four workloads (S1-S4) for each machine, by expanding the percentage of jobs requesting burst buffers to 50\% (S1 and S3 workloads) and 75\% (S2 and S4 workloads). In these synthetic workloads, the assigned burst buffer request is randomly selected from the original burst buffer requests in a certain range. S1 and S2 select requests from original requests greater than 5TB, while S3 and S4 choose from requests greater than 20TB. 
Figure \ref{workloads_pdf} shows the distributions of burst buffer requests for all the ten workloads (two systems, each with five workloads). As we can see, S3 and S4 workloads have larger burst buffer requests than S1 and S2. S1 and S2 have similar distributions, but more jobs in S2 request burst buffers. The similar pattern is observed in S3 and S4.

\subsection{Evaluation Metrics}\label{Evaluation Metrics}
There are two classes of metrics for evaluating job scheduling:  one is to evaluate system-level performance \textit{from the perspective of system managers}, and the other is to evaluate the quality of service \textit{from the perspective of users}.  In our experiments, we use four well-established metrics: 
\begin{itemize}
\item \textbf{Node usage} 
\footnote{Usage is a metric measuring resource utilization without considering availability.} 
measures the ratio of the used node-hours for useful job execution to the elapsed node-hours. 
\item \textbf{Burst buffer usage} measures the ratio of the used burst buffer hours to the elapsed burst buffer hours. 
\item \textbf{Job wait time} measures the interval between job submission to job start time. 
\item \textbf{Job slowdown} measures the ratio of the job response time (job runtime plus wait time) to its actual runtime. It is used to gauge the responsiveness of a system. We filter out abnormal jobs in calculating average slowdown, because many abnormal jobs end abruptly at beginning of execution leading to extremely high slowdowns.
\end{itemize}

Note that the first two metrics are system-level performance metrics, whereas the last two are user-level performance metrics.

In the rest of the paper, the 1st half month data is used to ``warm up'' the system and the last half month data is used to ``cool down'' the system. We present the results in the remaining months.

\subsection{Scheduling Methods}\label{Scheduling Methods}
We compare eight multi-resource scheduling methods:
\begin{itemize}
\item \textbf{Baseline}: Baseline represents the naive method for multi-resource scheduling (e.g., the one adopted in Slurm for burst buffer scheduling).
\item \textbf{Weighted}: This method aims to maximize a weighted combination of multiple objectives. The weights are site tunable parameters and system administrators can adjust them based on the importance of different resources. For this method, the weights of node utilization and burst buffer utilization are set to 50\% and 50\% respectively. These weights present the case where CPU and burst buffer are considered equally important.
\item \textbf{Weighted\_CPU}: In this weighted method, the weights of node utilization and burst buffer utilization are set to 80\% and 20\% respectively. These weights present the case where CPU is considered more important.
\item \textbf{Weighted\_BB}: In this weighted method, the weights of node utilization and burst buffer utilization are set to 20\% and 80\% respectively. These weights present the case where burst buffer is considered more important.
\item \textbf{Constrained\_CPU}: It aims to maximize node utilization under the constraints of burst buffers.
\item \textbf{Constrained\_BB}: It aims to maximize burst buffer utilization under the constraints of node utilization.
\item \textbf{Bin\_Packing}: This method is analogous to the bin packing method used in \cite{Grandl}. We compute alignment score (a dot product between the vector of machine's available resources and the job's requested resources) for jobs in the window and then allocate jobs with highest alignment score recursively until the machine cannot accommodate any further jobs.
\item \textbf{BBSched}: The scheduling scheme presented in this work. By default, we set the window size to 20, the number of generation to 500, the population size to 20, and the mutation probability to 0.05\%.
\end{itemize}

In our experiments, each of these multi-resource scheduling methods runs along with a base scheduler. With the Cori workloads, FCFS is used as the base scheduling method. With the Theta workloads, WFP (described in \Cref{HPC Job Scheduling}) is used as the base scheduling method. To make fair comparisons, we use the same window size for all methods. In addition, all the methods use EASY backfilling \cite{Feitelson02} to mitigate resource fragmentation.

\subsection{Results}\label{Results}

\begin{figure}[htp]
  \centering
  \subfigure{\includegraphics[scale=0.32]{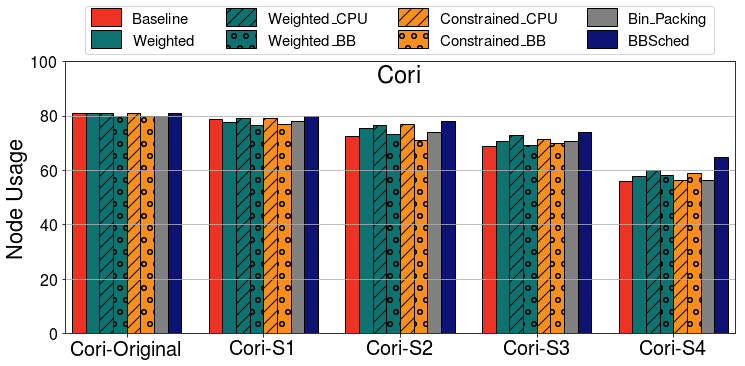}}
  \subfigure{\includegraphics[scale=0.32]{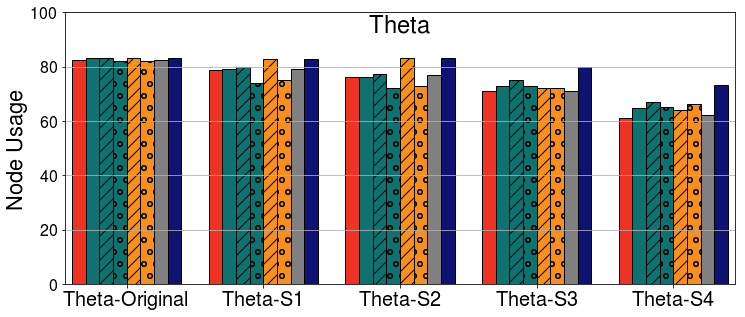}}
  \caption{Comparison of node usage on Cori traces (top) and Theta traces (bottom). The performance of the methods favoring CPU resource (Weighted\_CPU and Constrained\_CPU) are presented by the bars with lines on them, while the performance of the methods biasing towards burst buffers (Weighted\_BB and Constrained\_BB) are presented by the bars with cycles on them. }
  \label{node_util}
\end{figure}

\textbf{Impact on Node Usage:} Figure \ref{node_util} compares node usage obtained by different scheduling methods. Among all the methods, BBSched yields the best node usage for 7 out of 10 workloads. The most noticeable gains happened on Theta-S4 and Cori-S4. In the other 3 out of 10 workloads, namely Cori-Original, Theta-S1, and Theta-S2, Constrained\_CPU method achieves the best node usage. However, the performance difference between Constrained\_CPU and BBSched is negligible. When burst buffer is abundant, Constrained\_CPU method obtains good performance on node usage because it only optimizes node usage, whereas BBSched has to make trade-offs between node usage and burst buffer usage. When burst buffer becomes scarce, BBSched outperforms Constrained\_CPU method in terms of node usage. This is because burst buffer shortage has become the bottleneck of allocation of CPU resource to jobs with burst buffer requests. BBSched considers both resources and therefore eases the burst buffer contention between jobs and mitigates resource wastage on CPU. We also notice that Weighted\_BB and Constrained\_BB have very poor performance on node usage with the worst reductions in 6.11\% and 4.84\% respectively compared with the baseline. Considering that Weighted\_BB and Constrained\_BB favor burst buffers, they prioritize jobs that can make better use of burst buffers, and as a result, waste CPU resources.

\begin{figure}[h]
  \centering
  \subfigure{\includegraphics[scale=0.32]{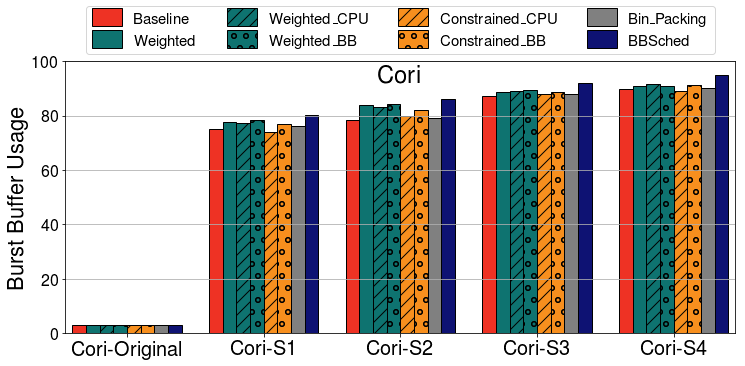}}
  \subfigure{\includegraphics[scale=0.32]{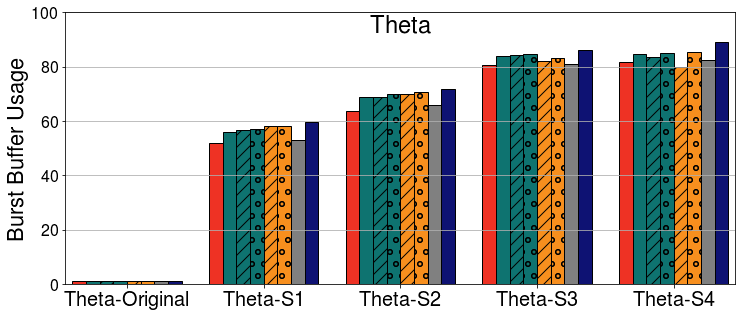}}
  \caption{Comparison of burst buffer usage on Cori traces (top) and Theta traces (bottom).}
  \label{BB_util}
\end{figure}

\textbf{Impact on Burst Buffer Usage:} Figure \ref{BB_util} shows burst buffer usage obtained by the eight scheduling methods. 
It is clear that all the methods except Constrained\_CPU improve burst buffer usage. The unsatisfactory performance obtained by Constrained\_CPU method is because this method puts node usage as its sole optimization objective and ignores burst buffer usage. This result clearly shows that exploiting job's complementary resource demands is crucial for multi-resource scheduling. 
Although BBSched, Weighted and Weighted\_BB consider both node and burst buffer utilizations in scheduling, weighted methods can only find one solution, while BBSched can produce multiple solutions. Therefore, BBSched is more likely to find a better solution from multiple solutions, leading to higher resource utilization. In summary, BBSched yields the best performance on burst buffer usage for all the workloads, and the performance improvement is as high as 15.46\% over the baseline. 

Node usage and burst buffer usage are correlated. Under the scenarios that burst buffer usages are under 80\% (Theta-Original, Theta-S1, Theta-S2, Cori-Original, and Cori-S1), optimization methods can keep the node usage around 80\%. But in the remaining scenarios with heavy burst buffer requests, we observe the noticeable drops in node usage in comparison to the original workloads. This phenomenon indicates that heavy burst buffer requests cause wastage in node resources. Among all the optimization methods, Constrained\_CPU suffers the most from the increase in burst buffer requests. The node usage drops more than 15\% from Cori-S3 to Cori-S4 and from Theta-S2 to Theta-S3. We attribute this to Constrained\_CPU that ignores burst buffer usage and is, therefore, incapable of ease burst buffer contention between jobs. In contrast, BBSched markedly improves node utilization on Theta-S3 (12.69\%), Theta-S4 (20.03\%) and Cori-S4 (16.28\%) compared to the baseline, and reduces the differences in node usage among workloads. This is because when workloads are shifting from node-bound to burst-buffer-bound, plenty of nodes will be left unused, which provides more room for optimization.
We also find that the biased optimization methods are effective in improving the usage of their favorite resource, but are at the risk of decreasing the usage of other resources. For example, Weighted\_BB and Constrained\_BB increase burst buffer usage by 10.54\% and 12.09\% at the cost of decreasing node usage by 6.11\% and 4.84\% respectively on Theta-S1. Similarly, Constrained\_CPU improves node usage by 0.18\%, but reduces burst buffer usage by 1.33\% on Cori-S1. In contrast, unbiased optimization methods, e.g., Weighted method, are more likely to improve utilization of both resources. In addition, although Bin\_Packing is capable of improving both node and burst buffer usage, the improvement is no more than 3.72\%, which is significantly less than the gains of the optimization methods. This is because Bin\_Packing selects jobs iteratively based on individual job information, but not attempts to find the best job combination which can optimize system performance. This demonstrates that optimization is necessary for improving system performance in multi-resource scheduling.

\begin{figure}[h]
  \centering
  \subfigure{\includegraphics[scale=0.32]{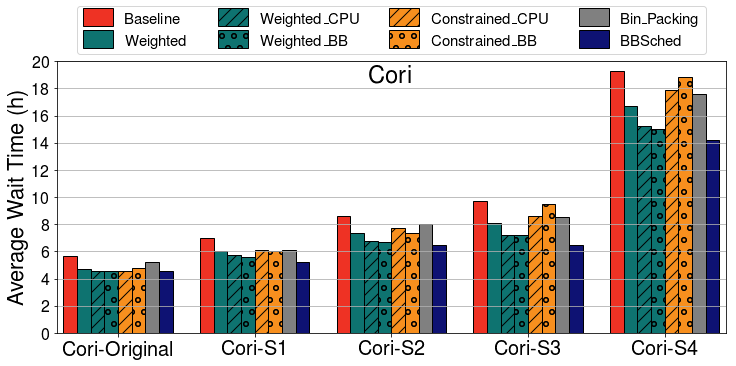}}
  \subfigure{\includegraphics[scale=0.32]{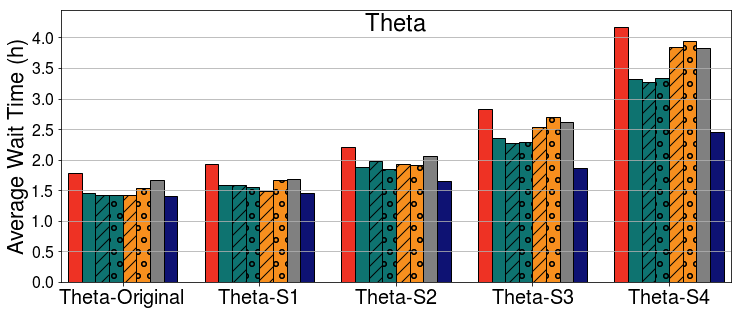}}
  \caption{Comparison of average job wait time on Cori traces (top) and Theta traces (bottom). The lower the average job wait time is, the better the performance is.}
  \label{awt}
\end{figure}

\textbf{Impact on Job Wait Time:} In Figure \ref{awt}, we compare average wait time of the eight scheduling methods. It is clear that all methods improve average job wait time in comparison to the baseline. BBSched achieves the most significant reductions on average job wait time, by up to 33.44\% on Cori and up to 41\% on Theta compared with the baseline. The second best method, Weighted\_BB, improves average job wait time by less than 26\% on both machines. We also notice that average job wait time increases dramatically as the burst buffer requests increases. For example, when using the baseline method, the average job wait time on Cori-Original is less than 6 hours, compared to 19 hours on Cori-S4. Additionally, the surge of burst buffer requests provides more opportunities for the optimization methods to reduce job wait time. For instance, BBSched reduces the average job wait time by 21.30\% on Theta-Original, while it reduces the average job wait time by 41\% on Theta-S4.

\begin{figure}[h]
\centering
\includegraphics[scale=0.32]{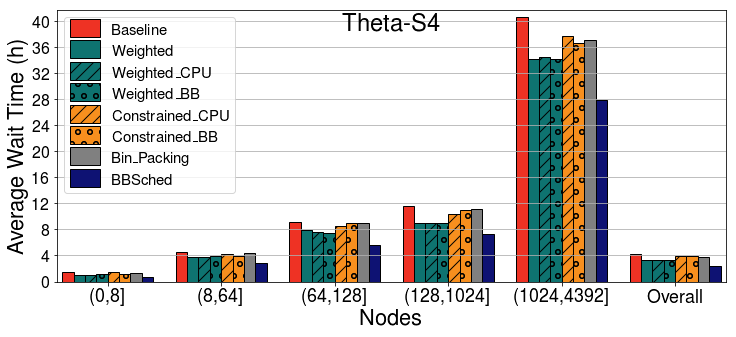}
\caption{Breakdown of average job wait times by job sizes on Theta-S4. }
\label{theta_awt_job_size}
\end{figure}

To understand the origin of the gains, Figure \ref{theta_awt_job_size} shows the breakdown of average job wait time by job sizes on Theta-S4. We observe that the most significant gain comes from small jobs. BBSched reduces average wait time by 48.29\% on 1-8 node jobs, compared to 31.59\% of reduction on 1024-4392 node jobs. The baseline method on Theta (WFP) prefers large jobs. However, small jobs on Theta wait less time than large jobs owing to EASY backfilling. The great wait time reductions on small jobs suggest that the optimization methods are more effective than EASY backfilling in avoiding resource fragmentation in multi-resource scheduling. We observe the similar results on all other workloads, and thus we only present the representative results on Theta-S4.

\begin{figure}[htp]
  \centering
  \subfigure{\includegraphics[scale=0.32]{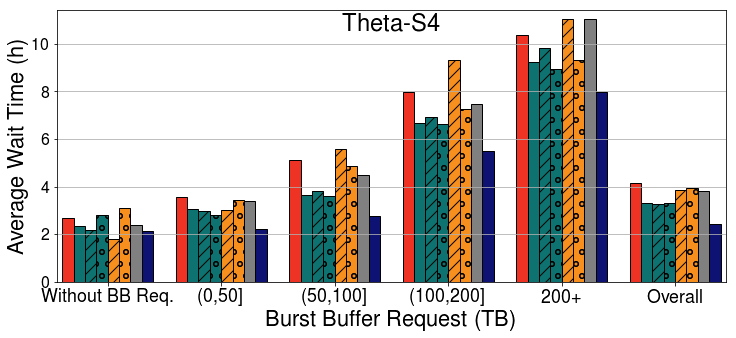}}
  \caption{Breakdown of average job wait times by burst buffer requests on Theta-S4. }
  \label{awt_BB}
\end{figure}

To show the impact of burst buffer requests, Figure \ref{awt_BB} presents the breakdown of average wait time by burst buffer requests on Theta-S4. Clearly, jobs with burst buffer requests wait longer times than jobs without burst buffer requests. For example, when using the baseline method, jobs with more than 200TB burst buffer requests wait, on average, 10.25 hours as opposed to 2.5 hours of jobs without burst buffer requests. We also observe that BBSched and weighted methods make more significant reductions on average jobs wait time of jobs with burst buffer requests. This is because they are designed to optimize both node and burst buffer utilization and therefore jobs with both node and burst buffer requests benefit more from these methods. In contrast, Constrained\_CPU fails to improve average wait time of jobs with burst buffer requests, because it only focuses on optimization CPU usage. For example, although Constrained\_CPU decreases the average wait time of jobs without burst buffer requests by 32.43\%, it increases the average wait time of jobs with 100-200TB burst buffer requests by 17.21\%. Under heavy burst buffer requests, such as Theta-S4, Constrained\_CPU optimizes node utilization by allocating jobs without burst buffer requests. But once jobs with burst buffer requests go over the threshold in the window, they are forced to run leaving a fraction of the nodes unused. Such inefficient job selections lead to poor performance on both node and burst buffer utilizations and job wait times. Constrained\_BB, on the hand, jeopardizes jobs without burst buffer requests to improve wait times of jobs with burst buffer requests. 

\begin{figure}[h]
\centering
\includegraphics[scale=0.32]{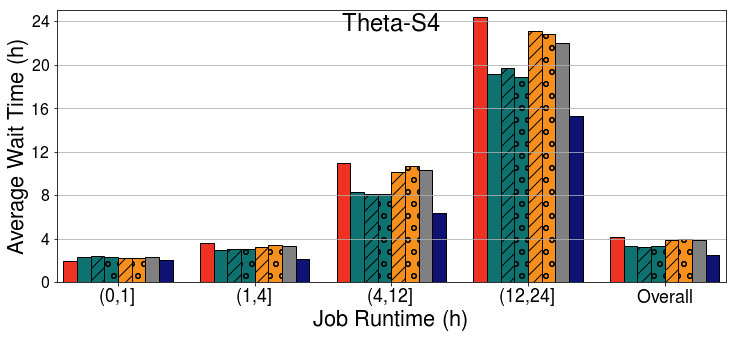}
\caption{Breakdown of average job wait times by job runtimes on Theta-S4. }
\label{theta_awt_job_length}
\end{figure}

\begin{figure}[h]
  \centering
  \subfigure{\includegraphics[scale=0.32]{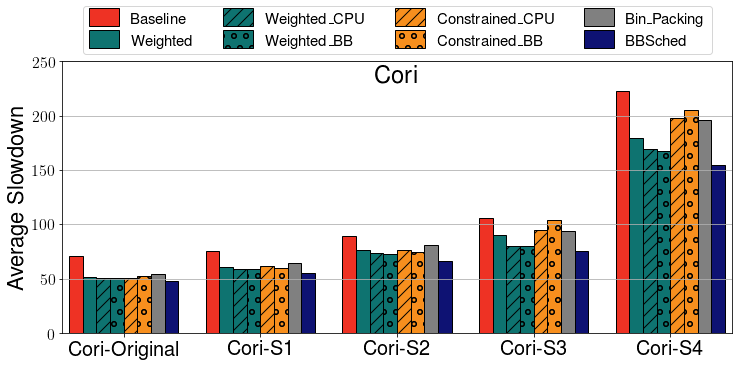}}
  \subfigure{\includegraphics[scale=0.32]{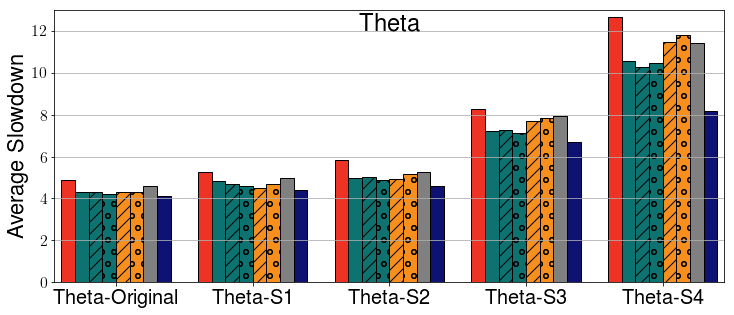}}
  \caption{Comparison of average slowdown on Cori traces (top) and Theta traces (bottom). The lower the average slowdown is, the better the performance is.}
  \label{Bslowdown}
\end{figure}

\begin{figure*}[h]
  \centering
  \subfigure{\includegraphics[scale=0.19]{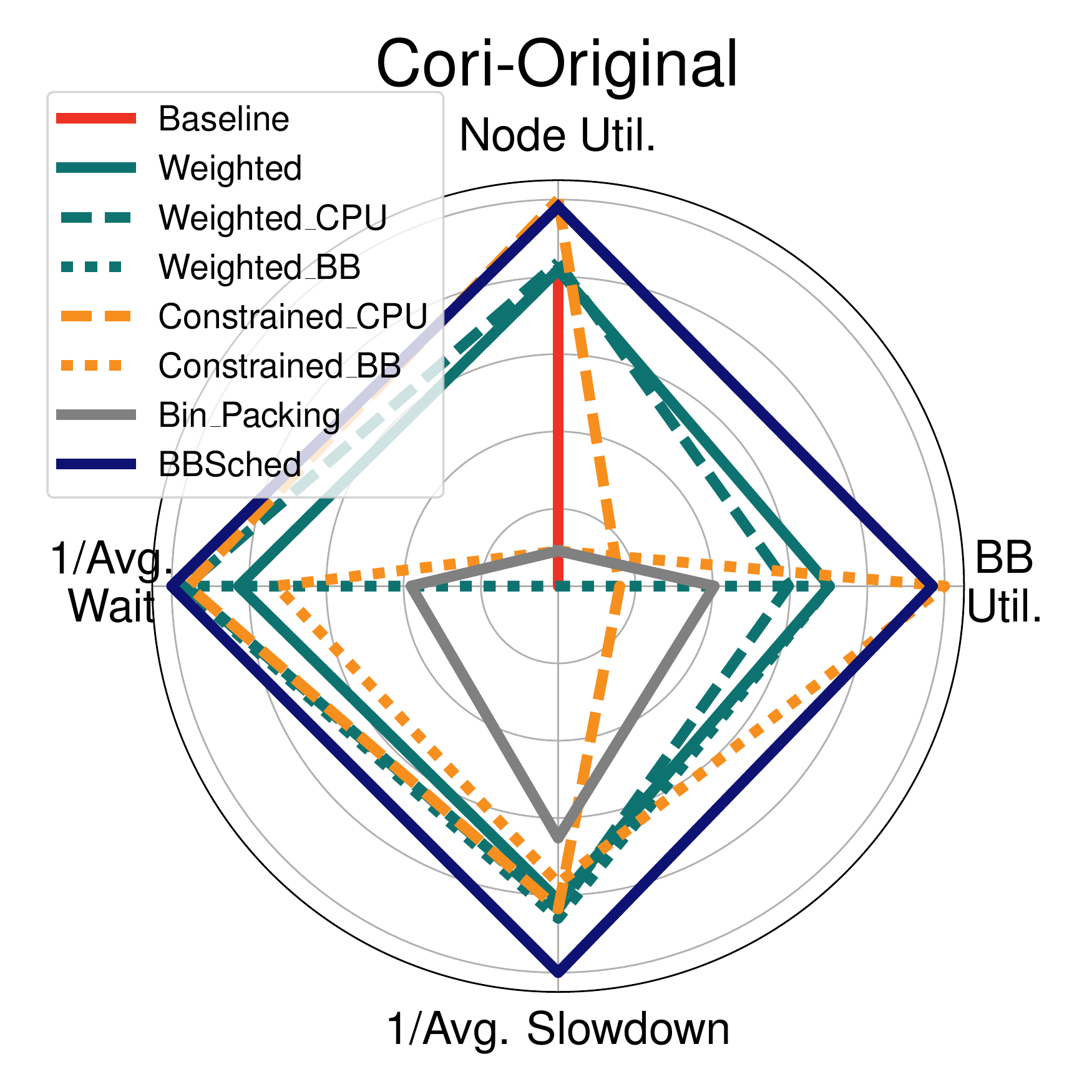}}
  \subfigure{\includegraphics[scale=0.19]{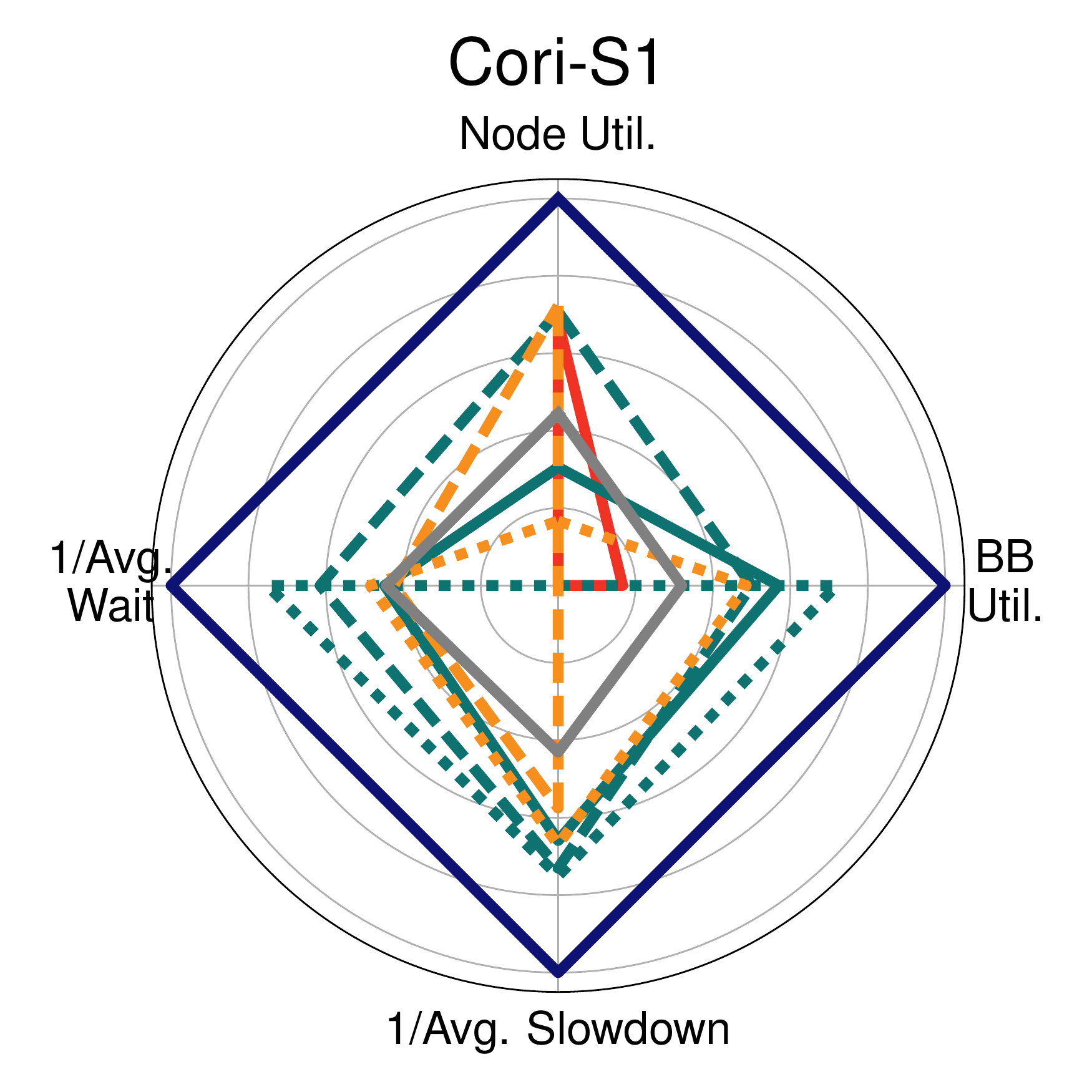}}
  \subfigure{\includegraphics[scale=0.19]{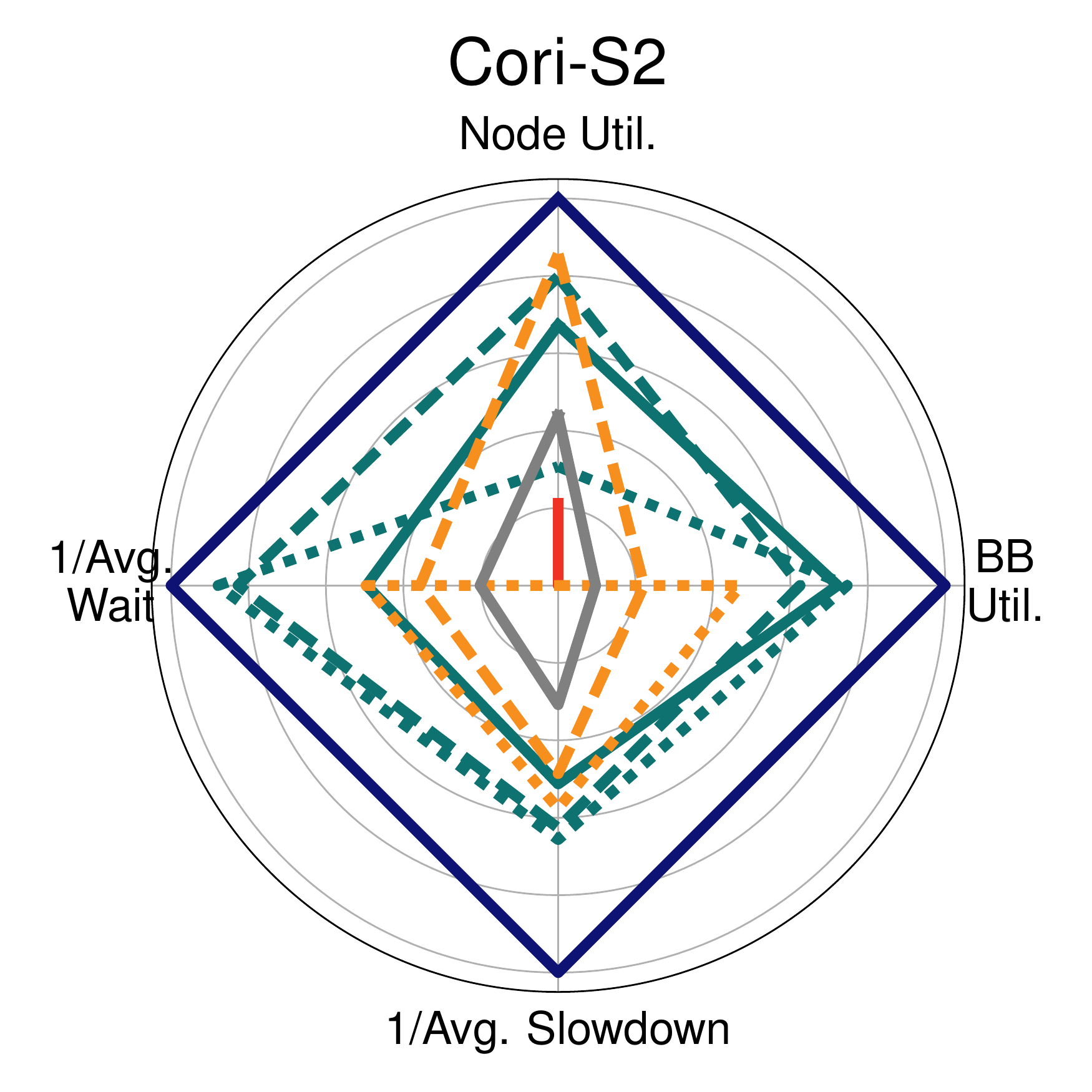}}
  \subfigure{\includegraphics[scale=0.19]{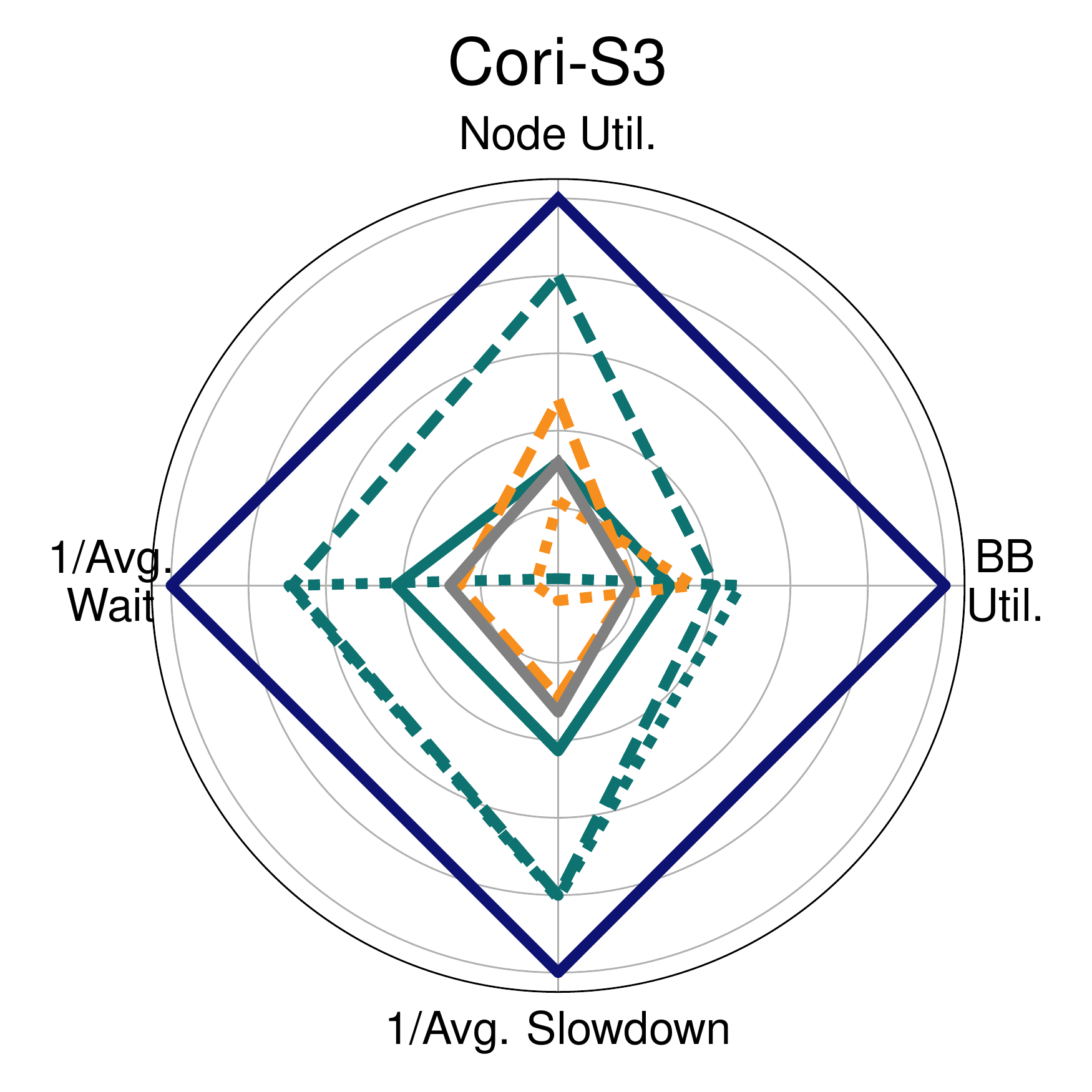}}
  \subfigure{\includegraphics[scale=0.19]{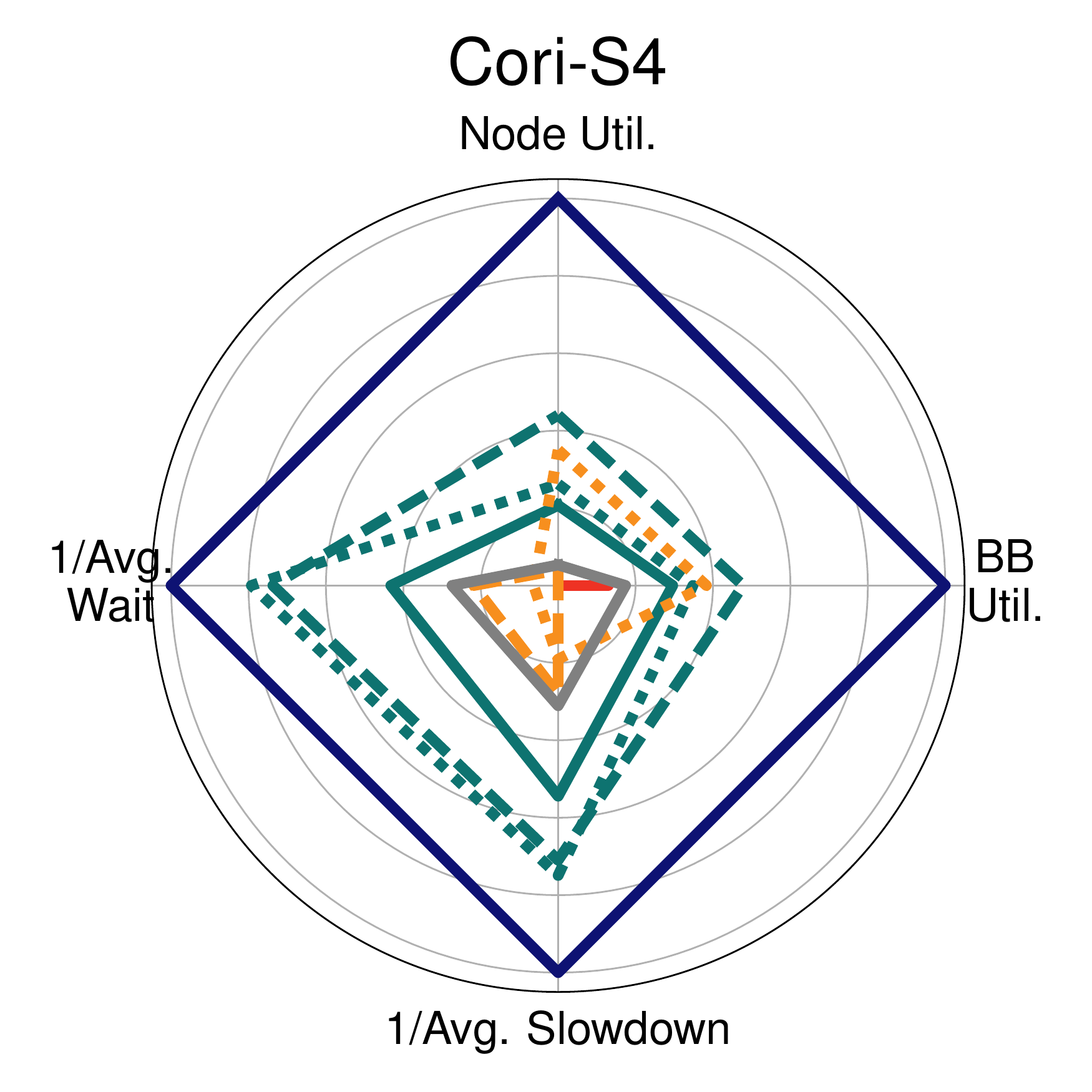}}
  \subfigure{\includegraphics[scale=0.19]{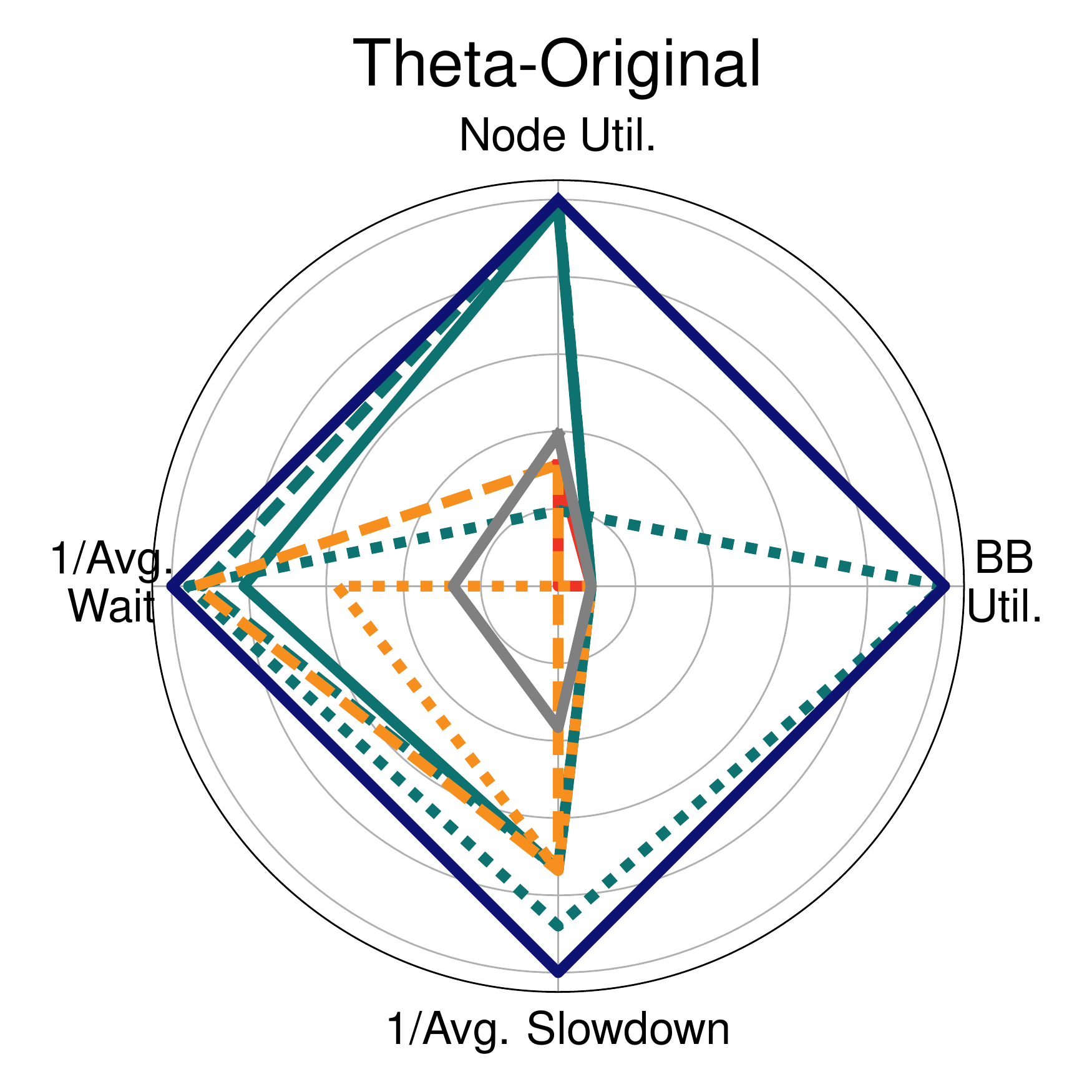}}
  \subfigure{\includegraphics[scale=0.19]{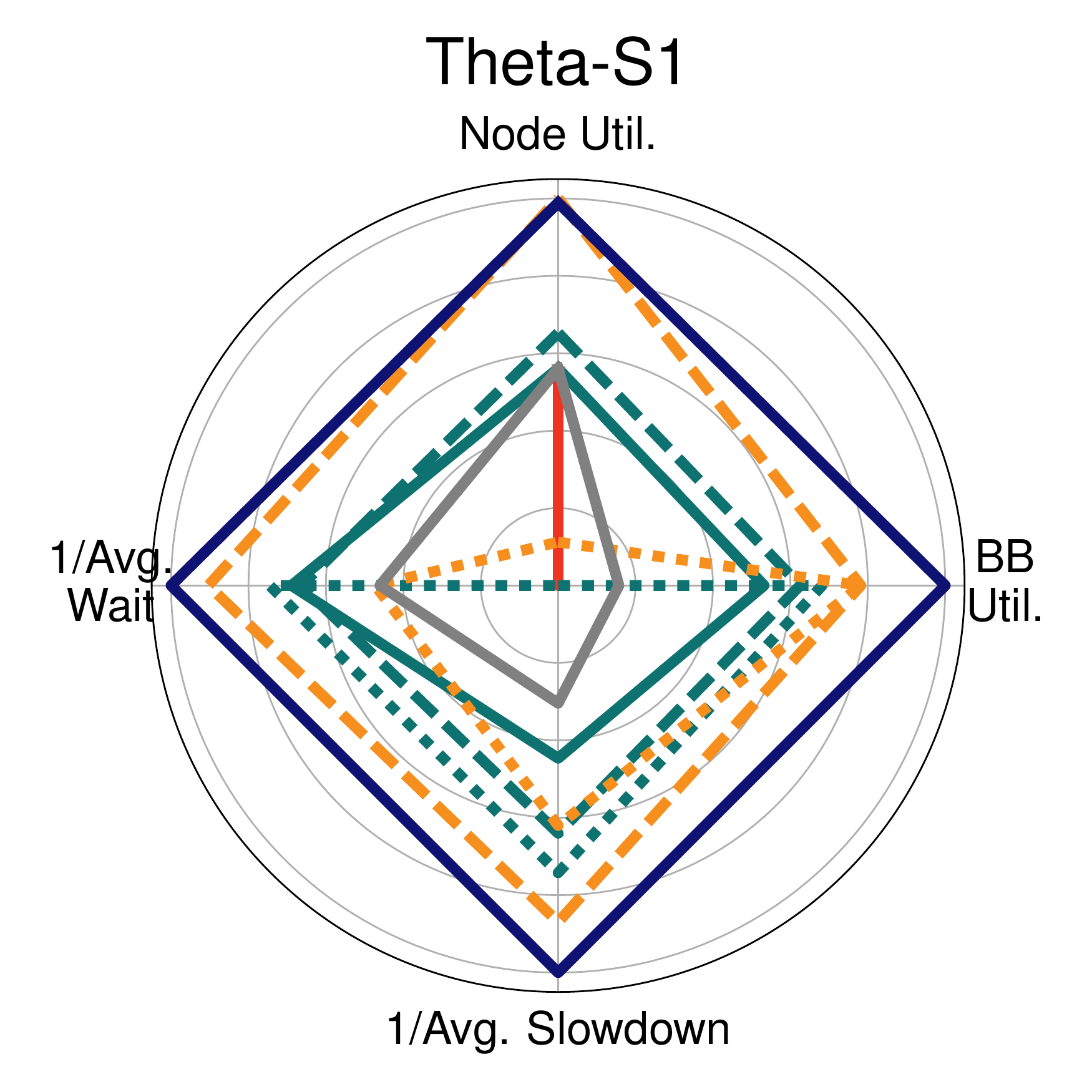}}
  \subfigure{\includegraphics[scale=0.19]{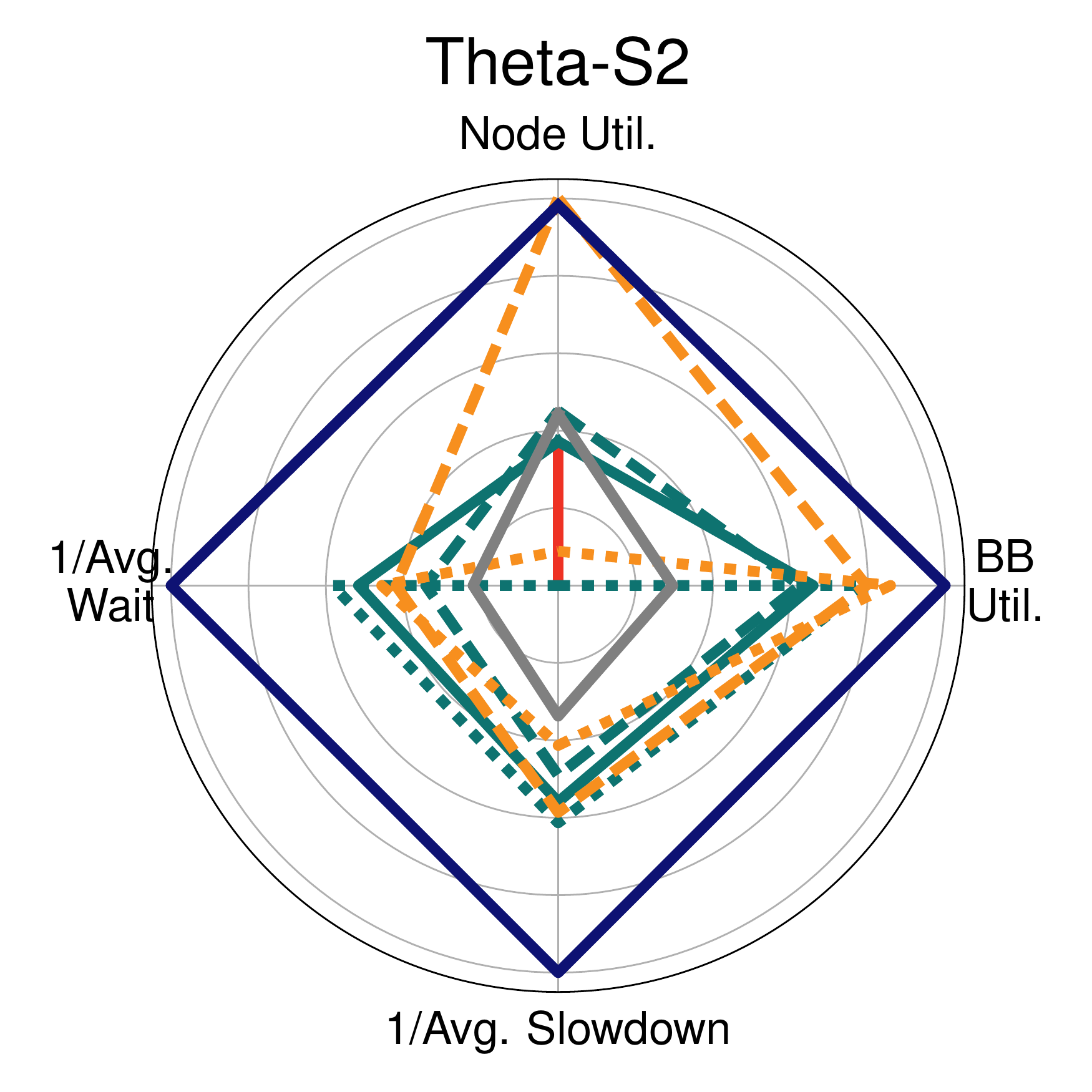}}
  \subfigure{\includegraphics[scale=0.19]{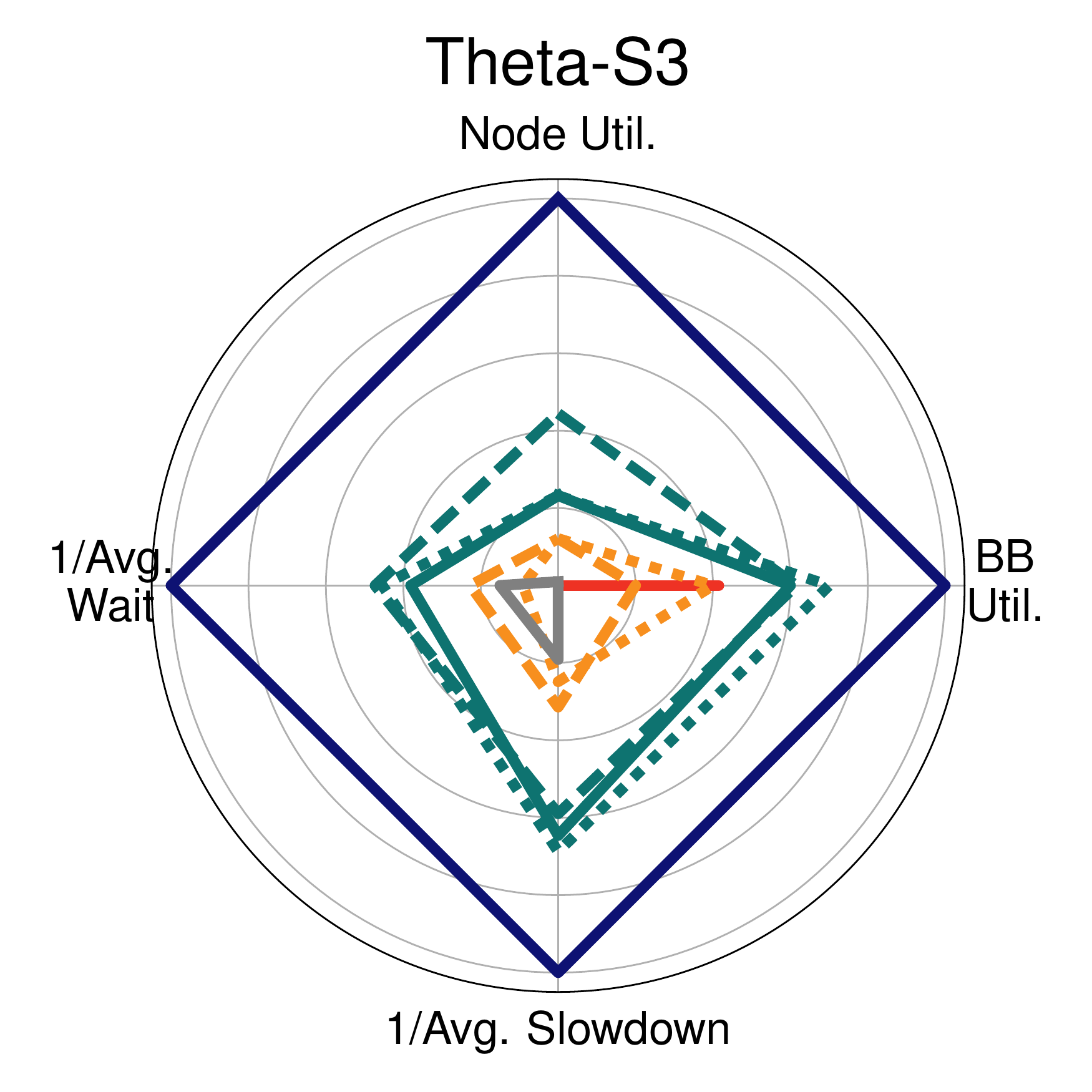}}
  \subfigure{\includegraphics[scale=0.19]{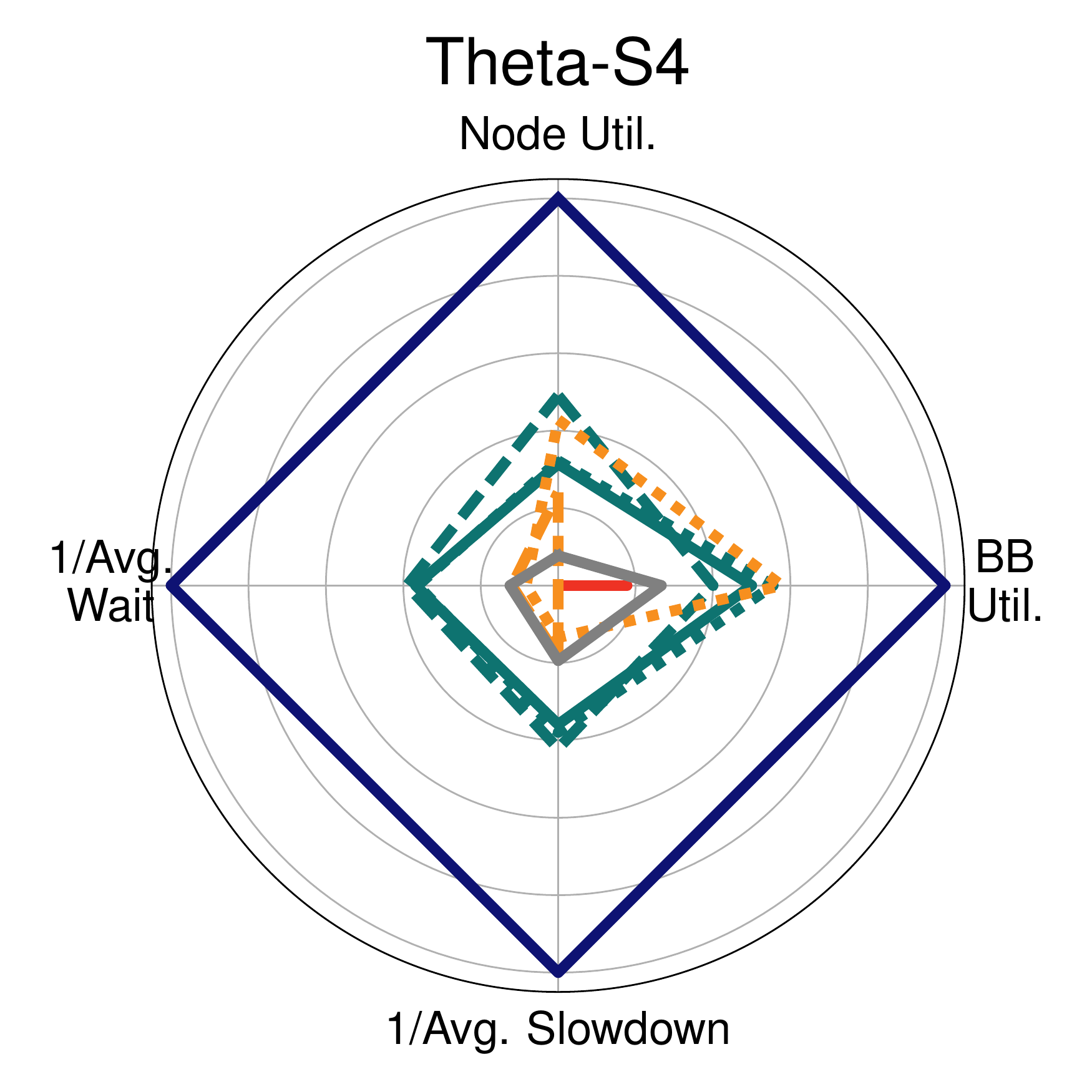}}
  \caption{Overall scheduling performance comparison using Kiviat graphs: Cori traces (top) and Theta traces (bottom). The larger the area is, the better the overall performance is.}
  \label{radar_chart}
\end{figure*}

If we look at the breakdown of average job wait time by job runtimes in Figure \ref{theta_awt_job_length}, we find that jobs wait times increase significantly with job runtimes. The reasons why short jobs wait less than long jobs are the baseline policy (WFP) and EASY backfilling. In WFP, shorter jobs get higher priorities to run. In EASY backfilling, jobs can be backfilled if it does not delay the first job in the queue. Hence, short jobs are more likely to be backfilled than long jobs. It is interesting to notice that all optimization methods reduce average wait time of long jobs, but increase the average wait time of short jobs. This is because the optimization methods only take job's resource requirement into consideration, which does not include job runtimes. As the optimization methods improves resource usage, less idle resources are left for backfilling. As a result, compared with the baseline, the optimization methods lead to longer wait times for short jobs owing to fewer opportunities for backfilling and shorter wait times for long jobs due to improvement in resource usage.

\textbf{Impact on Slowdown:} Figure \ref{Bslowdown} compares average slowdown of the scheduling methods. We find that the trends on average slowdown are similar to that on average wait time. Additionally, the performance of average slowdown is related to resource usage. The average slowdowns of Theta-S4 and Cori-S4 are evidently higher than other workloads. These workloads are also characterized by low node usage and high burst buffer usage. In these scenarios, even though more jobs are waiting to be scheduled, the severe burst buffer contention prevents jobs with burst buffer requests from being allocated, leaving a fraction of the nodes unused. 

\textbf{Holistic Performance Comparison:} To provide a holistic view of the performance of different methods, we present the scheduling results using Kiviat graph (see Figure \ref{radar_chart}). We use the reciprocal of average job wait time and the reciprocal of average slowdown in the plots. All metrics are normalized to the range of 0 to 1. $1$ means a method achieves the best performance among all methods and $0$ means a method obtains the worst performance. For all metrics, the larger the area is, the better the overall performance is.

Clearly, BBSched achieves the best and the most balanced performance, as it improves all the metrics significantly. Weighted methods and constrained methods make improvement on some metrics. Their overall performance, however, is unbalanced and is much lower than BBSched. Besides the baseline method, Bin\_Packing obtains the poorest performance, as it is an aggressive approach which makes scheduling decision based on isolated job information rather than information of multiple jobs. We also find that as the intensity of burst buffer requests increases, the areas of all methods except BBSched are shrinking. This suggests that BBSched can make notable performance improvement even under heavy burst buffer requests.

\begin{table}[]
\centering
\caption{BBSched performance under different window sizes. There are two numbers per cell: the top is for Cori-S4 and the bottom is for Theta-S4.}
\label{sen_window_size}
\begin{tabular}{l|lll}
\hline
 \backslashbox[30mm]{Metrics}{Window Size}                    & 10      & 20      & 50      \\ \hline
\multirow{2}{*}{CPU Usage}      & 60.18\% & 64.90\% & 65.06\% \\
                                      & 67.12\% & 73.29\% & 74.34\% \\ \hline
\multirow{2}{*}{Burst Buffer Usage} & 92.53\% & 94.74\% & 94.65\% \\
                                      & 84.23\% & 89.54\% & 89.63\% \\ \hline
\multirow{2}{*}{Average Job Wait Time (s)} & 55,732   & 51,028   & 50,871   \\
                                    & 10,402    & 8,847    & 8,792  \\ \hline
 \multirow{2}{*}{Average Slowdown} & 162.37   & 154.43   & 153.20   \\
                                    & 8.93    & 8.16     & 8.08  \\ \hline
\end{tabular}
\end{table}

\textbf{Sensitivity Analysis:}
Several parameters are used in the BBSched design. The selection of the number of generations ($G$) and population size ($P$) is discussed in \Cref{Burst Buffer Aware Scheduler}. The window size is another parameter in BBSched. Tuning the window size enables us to balance scheduling performance and preservation of the original job order. A larger window size leads to better resource utilization at the expense of higher computation overhead. Although window size does not directly affect computational complexity of BBSched (described in \Cref{Complexity Analysis}), a larger window size expands the search space and thus need more generations and larger population sizes to achieve acceptable approximation. Table \ref{sen_window_size} shows the sensitivity study on Cori-S4 and Theta-S4. As we can see, the most significant improvement is obtained when the window size is between 10 and 20. The improvement slows down with further increase in window size. Note that the Cori and the Theta workloads represent two types of HPC computing, namely capacity computing and capability computing. Considering that a larger window size can cause more disturbance to the original job order as well as higher computation overhead, we believe a window size of around 20 is an appropriate option for typical HPC workloads.

\textbf{Scheduling Overheads:} For methods other than BBSched, scheduling overhead depends on window size ($w$). As we increase $w$, the scheduling overhead increases. For BBSched, the number of generation ($G$) is the main factor that affects scheduling time. 

All experiments were performed on Intel Core i5 3.4GHz PC with 4 GB of RAM. It is not surprised that, besides the baseline, Bin\_Packing uses the least time (0.1s when $w$ is 50) in making scheduling decision. This is because Bin\_Packing is a greedy method rather than an optimization method. Although other methods take more time
, they still satisfy the time requirement of HPC scheduling (15-30 seconds). For example, if we set $G$ to 2000 and $w$ to 50 in BBSched, the average scheduling time is less than 2 seconds.

\section{Incorporating More Resources }\label{Extensibility}
The design of BBSched is generally applicable to schedule other shared or local resources. Local SSD is a representative local resource in HPC. Both Theta at ALCF and Summit at OLCF are equipped with local SSDs. On Theta, each node is equipped with a 128 GB local SSD and some will be gradually replaced by 256GB SSDs in the near future. In this section, we present a case study to illustrate that BBSched can be easily extended to incorporate additional schedulable resources. 

\textbf{Problem Formulation:}
Suppose a system has $N$ nodes and $B$ GB burst buffer. Each node is equipped with either 128GB SSD or 256GB SSD. $J = \{J_1,\dots,J_w\}$ is a set of $w$ jobs in the scheduling window: job $J_i$ requiring $n_i$ nodes, $b_i$ GB of shared burst buffer and $s_i$ GB of local SSD per node. For the $j$-th node assigned to job $J_i$, its actual local SSD volume $l_{ij}$ should be equal to or greater than the requested amount $s_i$. The difference between assigned SSD volume and requested SSD volume is considered as wasted local SSD volume.
In addition to the two objectives in \Cref{Burst Buffer Aware Scheduler}, we augment the MOO formulation with two additional objectives:

\begin{enumerate}
\item[(3)] maximize local SSD utilization: $f_3(\boldsymbol{x}) = \sum \limits_{i=1}^w s_i \times n_i  \times x_i$
\item[(4)] minimize wasted local SSD: $f_4(\boldsymbol{x}) = - \sum \limits_{i=1}^w (\sum\limits_{j=1}^{n_i}(l_{ij}-s_i)) \times x_i$
\end{enumerate} 
The multi-resouce scheduling problem can be formulated as:
\begin{align}
\text{max}  &  \quad  (f_1(\boldsymbol{x}),f_2(\boldsymbol{x}),f_3(\boldsymbol{x}),f_4(\boldsymbol{x})) \notag\\
\text{s.t.} &  \quad   \sum \limits_{i=1}^w n_i \times x_i \le N-N_{used},   \quad x_i \in \{0, 1\} \notag\\
&  \quad  \sum \limits_{i=1}^w b_i \times x_i \le B-B_{used},  \quad x_i \in \{0, 1\} \notag\\
&   \quad  s_i \le l_{ij},  \quad l_{ij} \in \{128, 256\} \notag
\end{align}
In the fourth objective, we minimize the wasted local SSD with the constraint that the assigned local SSD volume $l_{ij}$ is not less than the requested amount $s_i$.

\textbf{MOO Solver:}
The MOO solver basically remains the same and the only change is the rule of selecting the preferred solution in the decision maker. To accommodate one more resource, we adopt the following rule. First, we choose the solution with the maximum node utilization; next, we replace the preferred solution by another solution if the sum of the improvement in burst buffer utilization, local SSD utilization, and percentage of reduction in wasted local SSD of another solution is more than 4x of the loss of the node utilization. If we find more than one such solution, we choose the solution with the maximum sum of improvement.

\begin{figure}[h]
  \centering
  \subfigure{\includegraphics[scale=0.30]{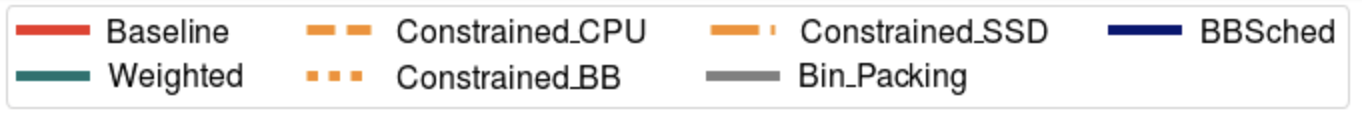}}
  \subfigure{\includegraphics[scale=0.23]{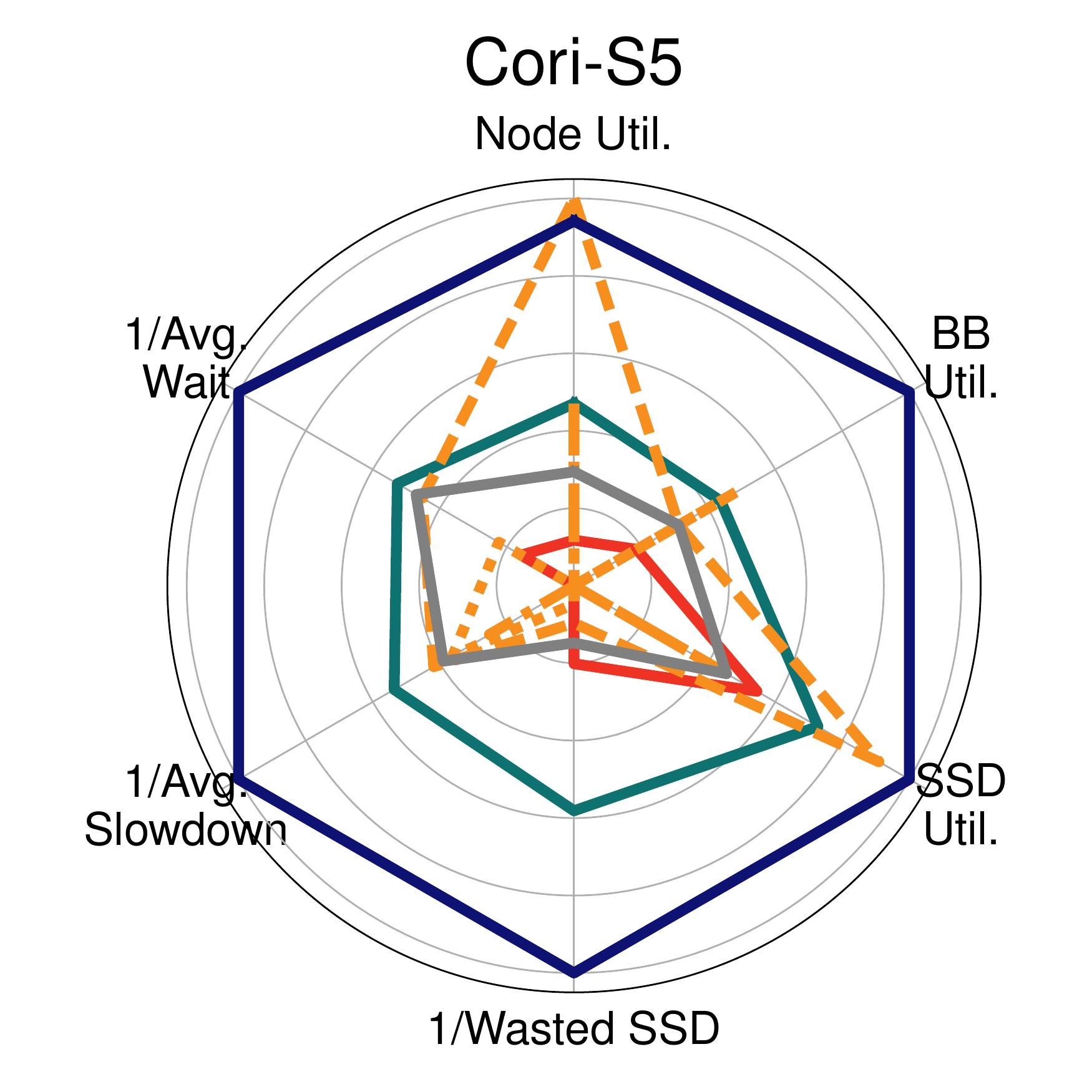}}
  \subfigure{\includegraphics[scale=0.23]{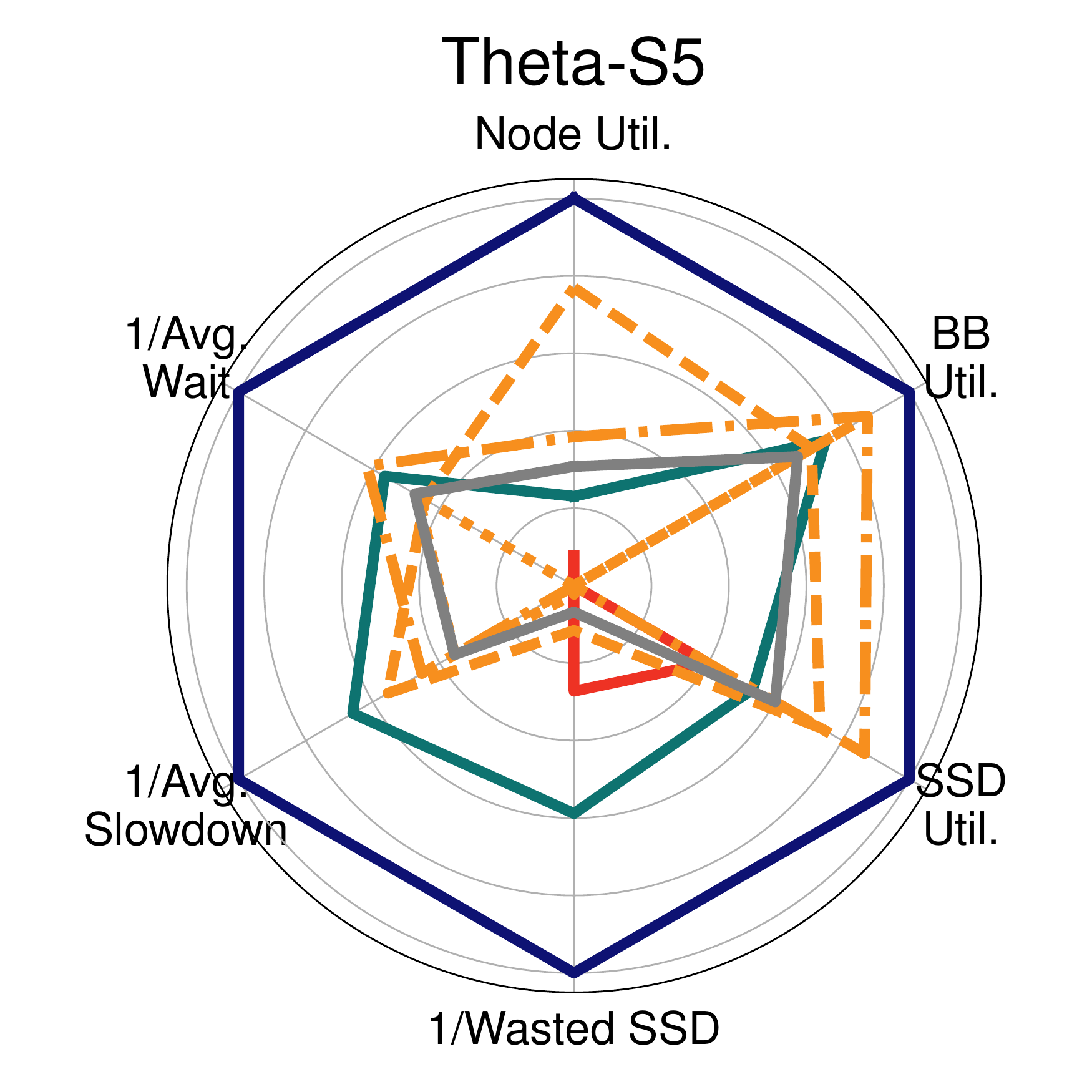}}
  \subfigure{\includegraphics[scale=0.23]{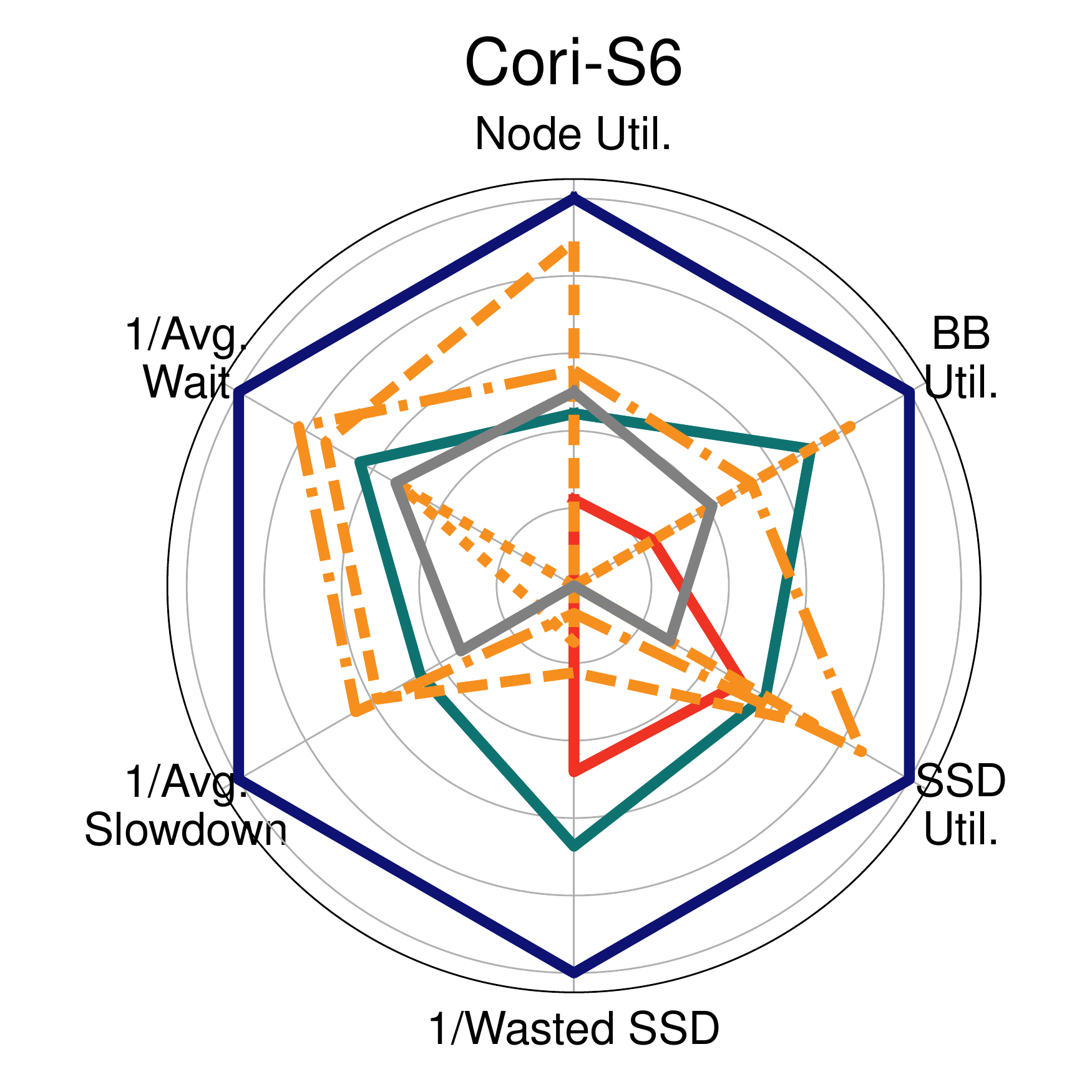}}
  \subfigure{\includegraphics[scale=0.23]{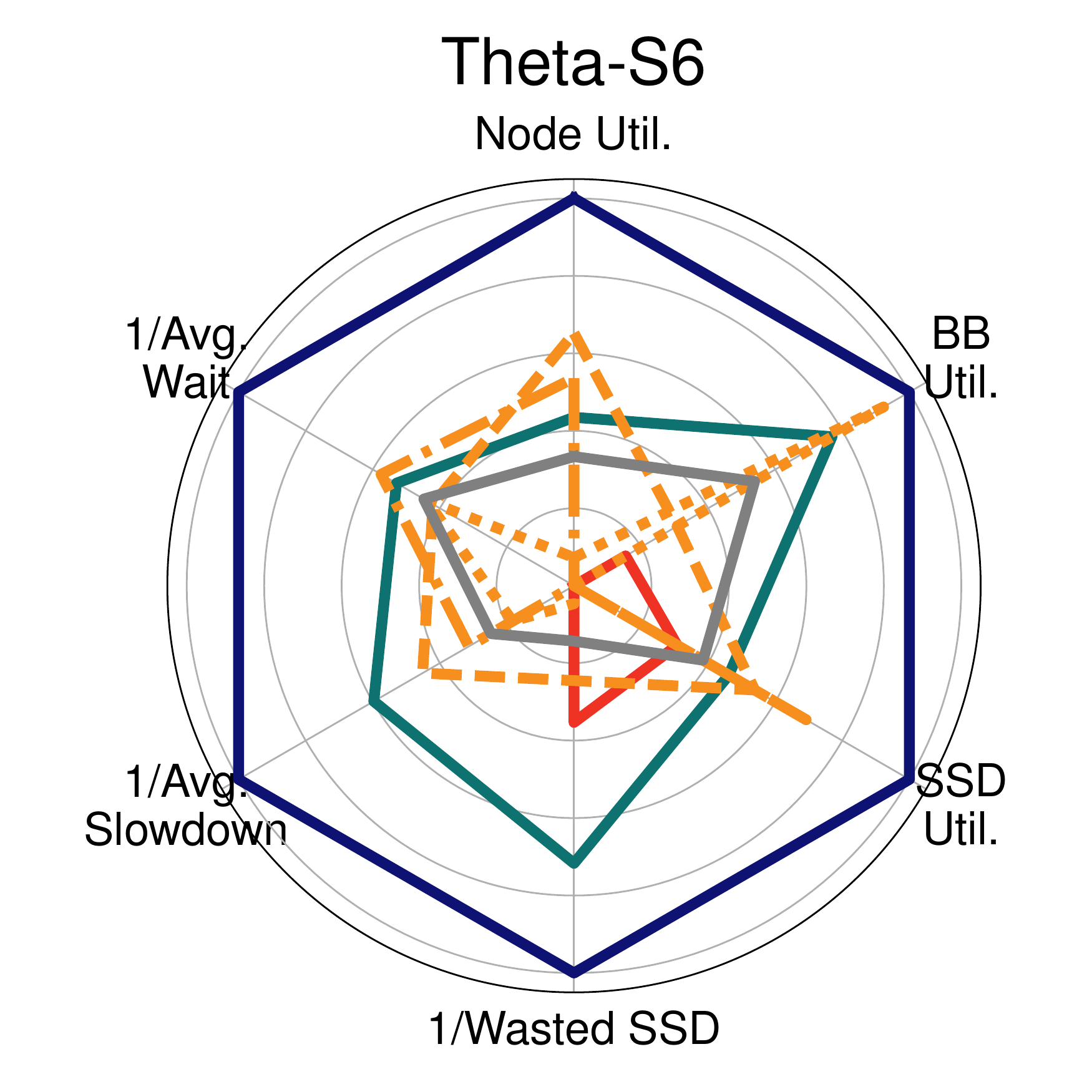}}
  \subfigure{\includegraphics[scale=0.23]{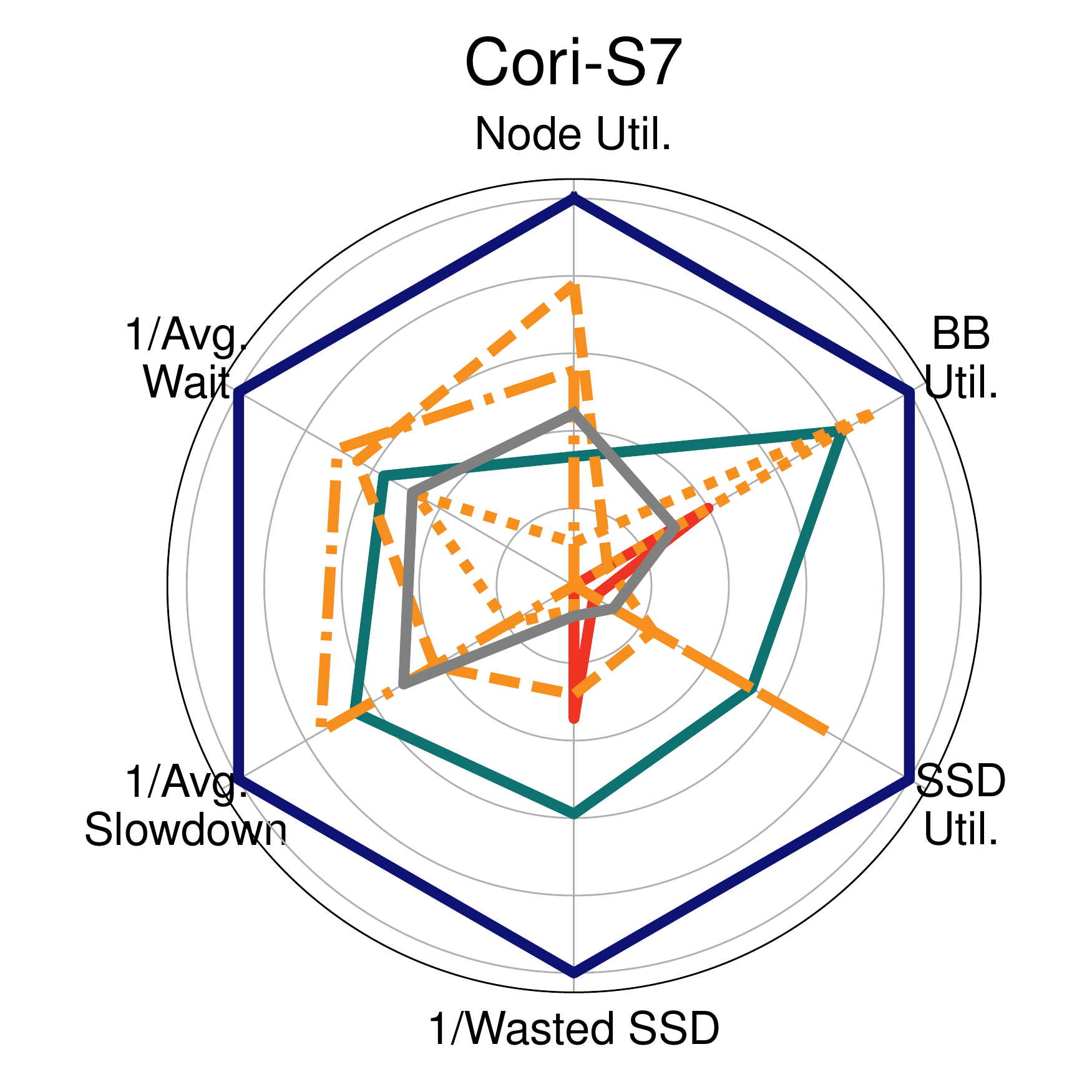}}
  \subfigure{\includegraphics[scale=0.23]{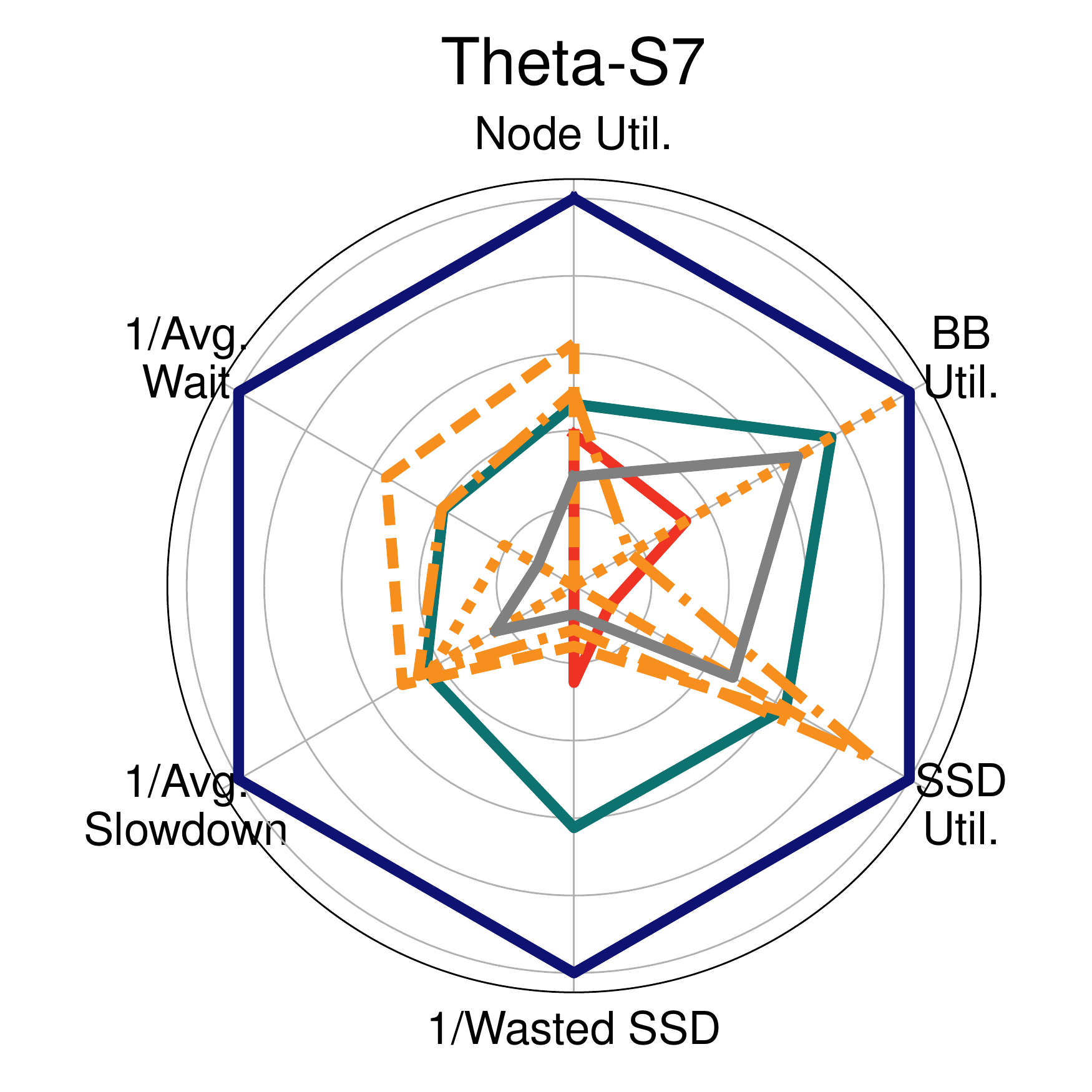}}
  \caption{Overall scheduling performance comparison using Kiviat graphs: Cori (left) and Theta (right). Note that as compared to Figure \ref{radar_chart}, the Kiviat graphs contain two additional metrics, that is, SSD utilization and the reciprocal of wasted SSD.}
  \label{radar_chart_three}
\end{figure}

\textbf{Workload Traces:}
In terms of the hardware configuration, we assume 50\% of nodes in the system are equipped with 128 GB local SSDs, the rest of the nodes are equipped with 256 GB local SSDs. We generate three workloads (S5-S7) on top of Cori-S2 and Theta-S2 by creating job's local SSD requests. In S5, 80\% of jobs have 0-128GB local SSD requests, and 20\% of jobs have 129-256GB local SSD requests. In S6, 50\% of jobs have 0-128GB local SSD requests, and 50\% of jobs have 129-256GB local SSD requests. In S7, 20\% of jobs have 0-128GB local SSD requests, and 80\% of jobs have 129-256GB local SSD requests. Jobs with more than 128GB local SSD requests have to be allocated to nodes with 256GB SSD. Jobs with no more than 128GB local SSD requests can be either allocated to nodes with 128GB or 256GB SSD. When assigning nodes to jobs with 0-128GB local SSD requests, nodes with 128GB SSD are preferred over 256GB SSD in order to mitigate wastage in local SSD. 

\textbf{Scheduling Methods:} We compare seven scheduling methods, i.e., Baseline, Weighted, Constrained\_CPU, Constrained\_BB, Constrained\_SSD, Bin\_Packing, and BBSched. Weighted method aims to maximize the equally weighted sum of node, burst buffer, local SSD utilization, and negative percentage of wasted SSD. Constrained\_SSD method aims to maximize local SSD utilization under the constraints of the other resources. The rest of the methods use the same strategies as described in \Cref{Scheduling Methods}.

\textbf{Results:}
Figure \ref{radar_chart_three} shows a holistic view of scheduling performance. We observe that BBSched achieves the best overall performance on all workloads. After BBSched, Constrained\_CPU and Constrained\_SSD methods have relatively good performance on both node utilization and local SSD utilization. This indicates that node utilization and local SSD utilization are correlated. Improving the utilization of one type of the resource increases the utilization of another resource. However, high local SSD utilization does not necessarily mean low wasted local SSD resource. The constrained methods and Bin\_Packing method waste local SSD resource more than the baseline, because they aggressively allocate nodes with 256GB SSD to jobs with 0-128GB SSD requests. Constrained\_BB obtains high burst buffer buffer utilization but low node and SSD utilization. Both Weighted and BBSched methods yield balanced performance. However, the improvement of Weighted method is noticeably lower than BBSched.

\section{Conclusion}\label{Conclusion}
In this paper, we have presented BBSched, a multi-resource scheduling scheme for HPC. 
In our design, the multi-resource scheduling problem is formulated into a MOO problem and is rapidly solved by a multi-objective genetic algorithm. BBSched generates a Pareto set for decision making, which enables system managers to explore potential tradeoffs among multiple resources for better utilization of multiple schedulable resources.

We have compared BBSched with existing methods using two real workloads and eight synthetic workloads. The extensive trace-based simulations demonstrate BBSched outperforms the existing methods in terms of both system-level and user-level metrics. Specifically, BBSched is capable of improving the overall performance by 41\% over naive method, 33\% over bin packing method, 35\% over constrained methods, and 20\% over weighted methods. \textit{This indicates that considering all resources in an explicit optimization is essential for HPC systems with multiple schedulable resources.} Moreover, we have presented a case study to show that BBSched can be extended to incorporate other resources (e.g., local SSDs in this case study). Given the promising results demonstrated in this study, our future work is to deploy and test BBSched on production systems.

\begin{acks}
This work is supported in part by US National Science Foundation grants CNS-1717763, CCF-1422009, CCF-1618776, and the U.S. Department of Energy, Office of Science, under contract DE-AC02-06CH11357, DE-AC02-05CH11231.
\end{acks}

\bibliographystyle{ACM-Reference-Format}
\balance
\bibliography{hpdc1.bib}

\end{document}